\documentclass[apj,twocolumn,twocolappendix,numberedappendix]{openjournal}
\usepackage{newtxtext,newtxmath,enumitem,multirow}
\usepackage[utf8]{inputenc}
\usepackage[T1]{fontenc}
\usepackage[breaklinks,colorlinks,allcolors=blue,citecolor=blue,urlcolor=blue]{hyperref}
\usepackage{orcidlink}
\usepackage{ctable}
\pdfoutput=1

\shorttitle{Dark matter content of UDGs}
\shortauthors{Kravtsov}

\begin{document}

\title[Dark matter content of UDGs]{On the dark matter content of ultra-diffuse galaxies\vspace{-1.5cm}}

\author{Andrey Kravtsov\,\orcidlink{0000-0003-4307-634X}$^{1,2,3,\star}$}
\affiliation{$^1$Department of Astronomy and Astrophysics, The University of Chicago, Chicago, IL 60637 USA}
\affiliation{$^2$Kavli Institute for Cosmological Physics, The University of Chicago, Chicago, IL 60637 USA}
\affiliation{$^3$Enrico Fermi Institute, The University of Chicago, Chicago, IL 60637}
\thanks{$^\star$\href{mailto:kravtsov@uchicago.edu}{kravtsov@uchicago.edu}}

\begin{abstract}
I compare the dark matter content within stellar half-mass radius expected in a $\Lambda$CDM-based galaxy formation model with existing observational estimates for the observed dwarf satellites of the Milky Way and ultra-diffuse galaxies (UDGs).
The model reproduces the main properties and scaling relations of dwarf galaxies, in particular their stellar mass-size relation. I show that the model also reproduces the relation between the dark matter mass within the half-mass radius, $M_{\rm dm}(<r_{1/2})$, and stellar mass exhibited by the observed dwarf galaxies. The scatter in the $M_{\rm dm}(<r_{1/2})-M_\star$ relation is driven primarily by the broad range of sizes of galaxies of a given stellar mass. I also show the $M_{\rm dm}(<r_{1/2})$ of UDGs are within the range expected in the model for their stellar mass, but they tend to lie above the median relation due to their large sizes. The upper limits on $M_{\rm dm}(<r_{1/2})$ for the dark matter deficient UDGs are also consistent with the range of dark matter masses expected in the model. The most dark matter-deficient galaxies of a given size correspond to halos with the smallest concentrations and the largest ratios of $M_\star/M_{\rm 200c}$. Conversely, the most dark matter-dominated galaxies are hosted by the highest concentration halos with the smallest $M_\star/M_{\rm 200c}$ ratios. 
Results presented in this paper indicate that dark matter content of observed UDGs may represent a tail of the expected dark matter profiles, especially if the effect of feedback on these profiles is taken into account. Nevertheless, the unusually rich GC populations in some UDG galaxies do indicate unusual evolution. Our results simply indicate that with the current accuracy of the velocity dispersion measurements their $M_{\rm dm}(<r_{1/2})$ are consistent with the range expected in the $\Lambda$CDM model. 
\end{abstract}


\maketitle

\section{Introduction}
\label{sec:intro}

The advent of deep imaging with CCD cameras resulted in discovery and investigations of the low surface brightness galaxies \citep[e.g.,][]{Bothun.etal.1985,Bothun.etal.1987,Impey.etal.1988,Schombert.etal.1992,Sprayberry.etal.1995,McGaugh.etal.1995,Dalcanton.etal.1997,Bothun.etal.1997,Impey.Bothun.1997,Driver.etal.2005,Geller.etal.2012}, confirming early conclusions of \citet{Disney.1976} that strong selection effects bias galaxy studies against low surface brightness objects. 

In the last decade hundreds of ultra-diffuse galaxies (or UDGs) were discovered in the outskirts of the Coma galaxy cluster  
\citep{vanDokkum.etal.2015,vanDokkum.etal.2015b,Koda.etal.2015,Yagi.etal.2016,Bautista.etal.2023} and other nearby clusters \citep{Conselice.etal.2003,van_der_Burg.etal.2016,van_der_Burg.etal.2017,Roman.Trugillo.2017,Venhola.etal.2017,Venhola.etal.2022,Mancera_Pina.etal.2018,Mancera_Pina.etal.2019,Gannon.etal.2022,Forbes.etal.2023,Zoeller.etal.2024,Marleau.etal.2024}, although indications that such galaxies are present in the Virgo cluster date back to \citet{Sandage.Bingelli.1984} and  \citet{Impey.etal.1988}. Ultra-diffuse galaxies have also been found in the field \citep[e.g.,][]{Leisman.etal.2017}. These gas-rich field UDGs are likely to be evolutionarily linked to dwarf irregular galaxies \citep{Nandi.etal.2023}. UDGs in clusters also represent a tail of the general dwarf spheroidal population \citep{Trujillo.2021,Nandi.etal.2023,Zoeller.etal.2024,Marleau.etal.2024}. 

Besides their large sizes and low surface brightnesses, many of these galaxies host surprisingly large globular cluster (GC) populations \citep{vanDokkum.etal.2017,vanDokkum.etal.2018b, Danieli.etal.2022}, although some nearby dwarf irregular galaxies of similar luminosities have comparably rich GC populations \citep{Georgiev.etal.2010,Forbes.etal.2018,Forbes.etal.2020}. 

Perhaps even more surprisingly, \citet{vanDokkum.etal.2018} reported velocity dispersion measurement in the UDG NGC1052-DF2 (hereafter DF2) and argued that the total mass estimate using this measurement is consistent with the stellar mass of the galaxy. This indicated that the galaxy is not dark matter dominated, as is usual for galaxies of similar luminosity. \citet{vanDokkum.etal.2019} reported a similar conclusion for the neighboring UDG NGC1052-DF4 (hereafter DF4) within the same group. Both of these galaxies have a sizeable globular cluster population with unusual GC luminosity function \citep{vanDokkum.etal.2018b}, and their GC luminosity function contains a strong peak at bright luminosities of $M_V\approx -9$ in addition to the usual peak at $M_V\sim -7.5$ \citep{Shen.etal.2021}. 

The conclusion about low dark matter content of DF2 and DF4 has been disputed. \citet{Martin.etal.2018} and \citet{Laporte.etal.2019} pointed out that uncertainties in velocity dispersion from globular clusters and luminosity measurement are substantial and the mass-to-light ratio for DF2 UDG is not constrained sufficiently well to exclude the presence of dark matter. They also argue that this galaxy is consistent with the scaling relations of Local Group dwarfs. 
\citet{Laporte.etal.2019} also pointed out that the dynamical mass estimated from velocity dispersion is a lower limit if there is significant rotation in the stellar or GC system. Indications of rotation have been reported for DF2 by \citet{Lewis.etal.2020}.

However, subsequent measurements presented by 
\citet{Danieli.etal.2019} and \citet{Shen.etal.2023} measured the velocity dispersion of DF2 and DF4 for stars and planetary nebulae, which indicated dispersion values consistent with the low values measured from globular clusters and thereby their low dark matter content \citep[see also][]{Emsellem.etal.2019}. They also did not find indications of significant rotation. 

Significant uncertainties and biases in the mass inference exist also \citep[e.g.,][]{Read.etal.2016b,Oman.etal.2019,Roper.etal.2023,Downing.Oman.2023,Sands.etal.2024} for the dark matter mass inference of gas-rich field UDGs based on the mass modeling of their HI velocity fields \citet{Guo.etal.2020,Mancera_Pina.etal.2020}. As I will discuss below, the mass modeling uncertainties from the rotation curve measurement for the Small Magellanic Cloud (SMC) allow the SMC to be considered a dark matter-deficient galaxy.  

As discussed in Section~\ref{sec:discussion} below, several scenarios have been proposed and discussed both for the origin of UDGs and the origin of their dark matter deficient subset. Formation scenarios for the latter often invoke special processes, such as galaxy collisions or formation out of gas stripped from galaxies as a result of tidal interactions or ram pressure stripping.  At the same time, some of the studies cited above argued that the UDG population, including their DM-deficient part, is shaped by standard galaxy formation processes such as stellar feedback and suppression of gas accretion in high-density regions. 

In this study, I re-visit the question of whether constraints on the dark matter content of the UDGs, including the dark matter deficient ones, are consistent with expectations of the galaxy formation in $\Lambda$CDM. Specifically, I use the galaxy formation model of \citet{Kravtsov.Manwadkar.2022}, which was demonstrated to reproduce properties of $L\lesssim L_\star$ galaxies down to ultra-faint dwarf luminosities at $z=0$ \citep{Kravtsov.Manwadkar.2022,Manwadkar.Kravtsov.2022,Kravtsov.Wu.2023}, as well as properties of galaxies at $z>5$ \citep{Kravtsov.Belokurov.2024,Wu.Kravtsov.2024}. Although some of the models discussed in Section~\ref{sec:discussion} attempt to explain both the dark matter content and properties of UDGs such as their sizes and GC population, I focus solely on their dark matter content in this study.

The paper is organized as follows. The samples of observed galaxies used for comparisons with model results, the halo samples used to compute the evolution of galaxies in our model, and the galaxy formation itself are presented in Section~\ref{sec:model}. The main results of this study are presented in Section~\ref{sec:results}, where it is shown that the dark matter content of UDGs is consistent with the expectations of the model and that UDGs follow the same relation between dark matter mass within the half-mass radius and stellar mass as the dwarf satellites of the Milky Way. These results are discussed in the context of previous studies of UDGs and their proposed formation scenarios in Section~\ref{sec:discussion}. The main conclusions of this study are summarized in Section~\ref{sec:conclusions}.

%-------------------
\section{Methods}
\label{sec:model}
%-------------------

%--------------------------------------------
\subsection{Observational galaxy sample}
\label{sec:obs_sample}
%---------------------------------------------

In this study galaxy formation model results will be compared to the observed dwarf satellites of the Milky Way using $V$-band luminosities, half-light radii, and velocity dispersion measurements in Table B1 of \citealt{Kravtsov.Wu.2023}, where measurements provenance is also specified. To convert $V$-band luminosities of satellites to stellar mass I use the $M_\star/L_V$ predicted by the galaxy formation model,  as quantified in \citet{Kravtsov.Manwadkar.2022}: $M_\star=(2.85 + 0.1M_V) L_V$, where $M_V$ is galaxy $V$-band absolute magnitude and its corresponding luminosity in solar luminosities is $L_V=10^{0.4(M_{V\odot}-M_V)}$ and $M_{V\odot}=4.81$ is the $V$-band absolute magnitude of the Sun \citep{Willmer.2018}.

In addition, I use a compilation of ultra-diffuse galaxies of \citet{Gannon.etal.2024cat}.\footnote{The compilation is available at \url{https://github.com/gannonjs/Published_Data/tree/main/UDG_Spectroscopic_Data}. Individual measurements are from \citet{mcconnachie2012,vanDokkum.etal.2015,Beasley2016,Martin2016,Yagi2016,MartinezDelgado2016,vanDokkum2016,vanDokkum.etal.2017,Karachentsev2017,vanDokkum.etal.2018,Toloba2018,Gu2018,Lim2018,RuizLara2018,Alabi2018,FerreMateu2018,Forbes2018,MartinNavarro2019,Chilingarian2019,Fensch2019,Danieli.etal.2019,vanDokkum.etal.2019b,Torrealba.etal.2019,Iodice.etal.2020,Collins.etal.2020,Muller2020,Gannon2020,Lim2020,Muller2021,Forbes2021,Shen.etal.2021b,Gannon2021,Gannon2022,Mihos2022,Danieli.etal.2022,Villaume2022,Webb2022,Saifollahi2022,Janssens.etal.2022,Gannon.etal.2023,FerreMateu.etal.2023,Toloba.etal.2023,Iodice.etal.2023,Shen.etal.2023}.} For consistency with the Milky Way dwarf galaxy sample, I use UDGs with stellar velocity dispersion measurements and estimate the total mass with stellar half-light radius using stellar line-of-sight velocity dispersion. 

The ultra-diffuse galaxies are distinguished by the number of globular clusters into GC-rich and GC-poor systems. The GC-rich systems have more than 20 globular clusters and are listed in Table 2 of \citet[][I exclude VCC1448 because it does not meet UDG definition]{Forbes.Gannon.2024}, the GC-poor UDGs are all other UDGs in the \citet{Gannon.etal.2024cat} compilation with measured stellar velocity dispersions. The stellar masses and effective radii used here are from this compilation. 

The  GC-rich and GC-poor UDGs are shown by the points of the same color but different types in the plots below. This distinction is made not to emphasize a specific difference between the two subpopulations but to indicate how they are distributed in different correlations shown in the figures, which may be of interest to researchers studying these subpopulations. As the figures below show, there are no clear differences in the considered correlations between GC-rich and GC-poor populations. Note, however, that this distinction does not correspond to the different specific frequency of GCs or to how typical or unusual the GC population is. 

In addition, I use measurements for two DM-deficient galaxies NGC1052-DF2 \citep{vanDokkum.etal.2018,Danieli.etal.2019} and NGC1052-DF4 \citep{vanDokkum.etal.2019,Shen.etal.2023}. Hereafter, the latter two galaxies will be referred to as DF2 and DF4 for brevity. For these galaxies I adopt stellar masses of $2\times 10^8\, M_\odot$ and $1.5\times 10^8\, M_\odot$, the projected circularized half-light radii of $2$ and $1.6$ kpc, and $M_{\rm tot}(<r_{1/2})$ of $1.3\pm 0.8\times 10^8\,M_\odot$ and $8^{+6}_{-4}\times 10^7\,M_\odot$, respectively \citep{Danieli.etal.2019,Shen.etal.2023}.

Finally, in the comparisons of the stellar mass-size relation of model and observed dwarf galaxies I use the sample of dwarf satellites in the ELVES survey \citep[see Table 9 in][]{Carlsten.etal.2022}. 

%---------------------------------------------------------
\subsection{Halo tracks and catalogs}
\label{sec:halocats}
%------------------------------------

I model a population of galaxies in halos of virial mass $M_{\rm 200c}\lesssim 10^{12}\, M_\odot$ using halo evolutionary tracks from several suites of high-resolution $N$-body simulations of zoom-in regions around MW-sized halos. Specifically, I use the Caterpillar \citep{Griffen.etal.2016} suite of $N$-body simulations\footnote{\url{https://www.caterpillarproject.org}} of 32 MW-sized haloes at the highest resolution level LX14 to maximize the dynamic range of halo masses probed by our modelling. The halo masses resolved in these simulations is sufficient to model the full range of luminosities of observed Milky Way satellites, as faintest ultrafaint dwarfs are hosted in haloes of $M_{\rm peak}\gtrsim 10^7\, M_\odot$ in our model \citep[][]{Manwadkar.Kravtsov.2022}. The Caterpillar suite was simulated assuming flat $\Lambda$CDM cosmology with the Hubble constant of $h=H_0/100=0.6711$, the mean dimensionless matter density of $\Omega_{\rm m0}=0.32$, the rms fluctuations of matter density at the $8h^{-1}$ Mpc scale of $\sigma_8=0.8344$ and primordial power spectrum slope  $n_s=0.9624$. 

In addition, I use halo tracks and catalogs from the ELVIS \citep{Garrison_Kimmel.etal.2014} and Phat ELVIS \citep[with central disk and without][]{Kelley.etal.2019} suites of zoom-in simulations of MW-sized halos. The mass and force resolution of these simulations are somewhat lower than those of the Caterpillar simulations (except for the two HiRes halos in the ELVIS suite). I use halos of mass $M_{\rm 200c}>10^8\, M_\odot$ from these simulations to complement halos from the Caterpillar suite to improve statistics of dwarf galaxies with stellar masses $M_\star>10^5\, M_\odot$. The ELVIS simulations were run assuming flat $\Lambda$CDM model with $h = H_0/100=0.71$, $\Omega_{\rm m0} = 0.266$,  $\sigma_8 = 0.801$, and $n_s=0.963$,  while Phat ELVIS simulations assumed $h=0.6751$, $\Omega_{\rm m0}= 0.3121$, $\sigma_8=0.81$, and $n_s=0.968$. 

I use evolution tracks of halos and subhalos that exist at $z=0$ in and around the Milky Way-sized halo within zoom-in region in these simulation suites to construct mass evolution tracks for the galaxy formation model described below. The tracks were extracted from simulations using the Consistent Trees code \citep{Behroozi.etal.2013} and consist of  several halo properties, such as its virial mass, scale radius, maximum circular velocity, etc., measured at a series of redshifts from the first epoch at which progenitors are identified to $z=0$. I use properties of dark matter halos at $z=0$ to model the mass distribution in these halos. 

Tidal stripping may affect dark matter content of some of the UDGs by removing dark matter from the inner regions of the galaxy. Nevertheless, in this analysis, to gauge whether UDG properties are consistent with the expectations for galaxies in isolated halos I compare observed UDGs to the population of galaxies in halos that are not strongly affected by tidal stripping. To this end, I only consider halos with the ratio of the current mass with the peak virial mass along the track of $M_{\rm 200c}/M_{\rm 200c,peak}>0.5$.

%---------------------------------------------------------
\subsection{Galaxy formation model}
\label{sec:galform_model}
%------------------------------------

The mass evolution tracks of halos in the zoom-in regions of the simulations described above were used as input for the \texttt{GRUMPY} galaxy formation model  \citep{Kravtsov.Manwadkar.2022}. This is regulator-type galaxy formation model \citep[e.g.,][]{Lilly.etal.2013} based on the models of \citet{Krumholz.Dekel.2012} and \citet{Feldmann.2013}, but with a number of modifications to extend the model into the dwarf galaxy regime. The model 
evolves key properties of gas and stars (masses, metallicities, sizes, star formation rates, etc.) of the galaxies they host by solving a system of coupled differential equations governing the evolution of these properties. The model accounts for the UV heating after reionization and associated gas accretion suppression onto small mass haloes, galactic outflows, a model for gaseous disk and its size, molecular hydrogen mass, star formation, etc.  The evolution of the half-mass radius of the stellar distribution is also modeled. The galaxy model parameters used in this study are identical to those used in \citet{Manwadkar.Kravtsov.2022}. 

The \texttt{GRUMPY} model is described and tested against a wide range of observations of local dwarf galaxies in \citet[][]{Kravtsov.Manwadkar.2022}. The model was shown to reproduce luminosity function and radial distribution of the Milky Way satellites and size-luminosity relation of observed dwarf galaxies \citep[][]{Manwadkar.Kravtsov.2022}.

\begin{figure}
  \centering
  \includegraphics[width=0.499\textwidth]{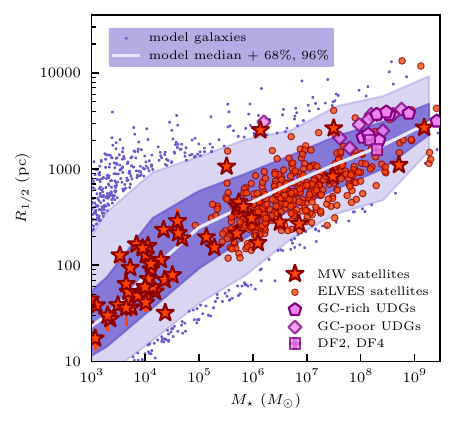}
  \caption[]{Half-mass radius of observed and model galaxies as a function of their stellar mass.  Observed galaxies (see Sec.~\ref{sec:obs_sample} for details and references) include Milky Way satellites ({\it red stars}), dwarf satellites in the ELVES host galaxy sample ({\it small circles}), GC-poor ({\it diamonds}) and GC-rich ({\it pentagons}) ultra-diffuse galaxies,  and two DM-deficient galaxies NGC1052-DF2 and NGC1052-DF4. The median half-mass radius in bins of stellar mass for model galaxies is shown by the white line, while shaded bands show the 16th and 96th percentiles of the distribution around the median; blue dots show individual galaxies outside these ranges.  The figure shows that the model reproduces $M_\star -R_{1/2}$ relation of observed galaxies and its scatter.}
   \label{fig:mstar_rhalf}
\end{figure}

Figure~\ref{fig:mstar_rhalf} shows the relation between galaxy's stellar mass and its stellar half-mass radius for model galaxies and observed galaxies. The latter include Milky Way satellites, the satellite galaxies in the ELVES sample \citep{Carlsten.etal.2022}, and the ultra-diffuse galaxies with existing measurements of stellar velocity dispersion (see Section~\ref{sec:obs_sample} for details and references). 
Figure~\ref{fig:mstar_rhalf} shows that the median galaxy size--stellar mass relation and the scatter around it are in good agreement with the relation of observed dwarf galaxies \citep[cf. also][]{Manwadkar.Kravtsov.2022}. 

Note that observational size estimates are for the projected half-light radii $R_{1/2}$ while in the model we follow 3d half-mass radii $r_{1/2}$. We do not explore the potential difference between these radii because conversion of $R_{1/2}$ to $r_{1/2}$ for observed galaxies is uncertain because it depends on the star formation history of galaxies \citep{Somerville.etal.2018,Suess.etal.2019} and ellipticity and radial density distribution of stars \citep{Somerville.etal.2018,Behroozi.etal.2022}. The factor $\chi$ relating the two radii $r_{1/2}=\chi R_{1/2}$ is thus expected to vary between $\chi\approx 0.85-1$ for disk systems to $\chi\approx 1.34-1.6$ for spheroidal systems.  However, information to estimate $\chi$ reliably is lacking for most of the individual galaxies. I therefore 
chose to keep $\chi=1$ for this analysis. 

Predictably, given their definition, UDGs have sizes close to or above the median of the relation. I can also note that the DM-deficient galaxies DF2 and DF4 have some of the smallest sizes among the UDG sample shown in this figure. 
The UDG with $M_\star\approx 1.6\times 10^6\, M_\odot$ with a half-light radius just above the 96\% band (farthest from the median among UDGs) is M31 dwarf spheroidal satellite And XIX \citep{Martin2016}. The Antlia II satellite of the Milky Way \citep{Torrealba.etal.2019} is located near this galaxy. There are also several outliers in the ELVES satellite sample of \citet{Carlsten.etal.2022}. This illustrates that ultra-diffuse galaxies exist and are fairly common among dwarf satellites of $L_\star$ galaxies. 

Note that the stellar mass-size relation is quite shallow at $M_\star>10^5\,M_\odot$, where size changes only by an order of magnitude over four orders of magnitude of stellar mass. The ultra-diffuse galaxies, which have $R_{1/2}>1.5$ kpc by definition, thus span a wide range of stellar masses and, likely, halo masses. This is consistent with the finding of \citet{Danieli.vanDokkum.2019} that UDGs with a given size range span $\approx 6$ absolute magnitudes or $\approx 2.5$ orders of magnitude in luminosity. 

In what follows, I use stellar masses and half-mass sizes produced by the model along with the information about halo mass and concentration of their parent halos from the halo catalogs, to estimate total and dark matter mass within the stellar half-mass radius, as I describe in the next section.

% %

%---------------------------------------------------------
\subsection{Mass profiles of model galaxies}
\label{sec:mhalf_model}
%------------------------------------

To estimate total masses $M_{\rm tot}(<r_{1/2})$ for model galaxies, I use individual half-mass radii, $r_{1/2}$, and consider two assumptions for the density profile of dark matter. Specifically, in the first model I adopt the Navarro-Frenk-White density profile \citep[NFW,][]{Navarro.etal.1997} for dark matter profile and use $M_{\rm 200c}$, $R_{\rm 200c}$, and the NFW scale radius, $r_s$, available in the halo catalogs:\footnote{Note that for subhalos the scale radius and $M_{\rm 200c}$ are estimated using only bound dark matter particles by the Rockstar halo finder \citep[][]{Behroozi.etal.2013}. }
\begin{equation}
    M_{\rm tot,NFW}(<r_{1/2}) = M_{\rm dm,NFW}(<r_{1/2}) + \frac{1}{2}\,M_\star
\label{eq:mhalfnfw}
\end{equation}
where $M_\star$ is the stellar mass of the galaxy hosted by a given halo computed in the model and 
\begin{equation}
    M_{\rm dm,NFW}(<r) = M_{\rm 200c}\,\frac{f(r/r_s)}{f(R_{\rm 200c}/r_s)};
\label{eq:mdmnfw}
\end{equation}
where
\begin{equation}   
    f(x) = \ln(1+x) - \frac{x}{x+1}.
\end{equation}

In the second model, I assume the dark matter density profile of \citet{Lazar.etal.2020}, which approximates the effects of stellar feedback on the density profile in the FIRE-2 simulations: 
\begin{equation}
    M_{\rm tot, L}(<r_{1/2}) = M_{\rm dm,Lazar}(<r_{1/2}) + \frac{1}{2}\,M_\star.
\label{eq:mhalflazar}
\end{equation}
Specifically, I use the parametrization of the cored-Einasto density profile in the equations~8-10, 12 of \citet{Lazar.etal.2020} and equations for the cumulative mass profile in their Appendix B1 and parameters in the second row of their Table 1 for the dependence of profile as a function of stellar mass $M_\star$. I chose to use the dependence on the stellar mass, to minimize the effects of different $M_\star/M_{\rm h}$ in their simulations and our model. Note that the profiles were calibrated only for galaxies of $M_\star\gtrsim 10^5\,M_\odot$. However, in this model effects of feedback for $M_\star<10^6\, M_\odot$ are expected to be negligible and thus extrapolating their results to smaller masses is equivalent to simply assuming Einasto profile with negligible core for these low-mass systems. 

Note that I neglect the gas mass in the equations~\ref{eq:mhalfnfw} and \ref{eq:mhalflazar}. Given that most of the observed UDGs I compare with are located in galaxy groups and clusters I assume that most of their ISM gas is stripped.  

% %
\begin{figure}
  \centering
  \includegraphics[width=0.475\textwidth]{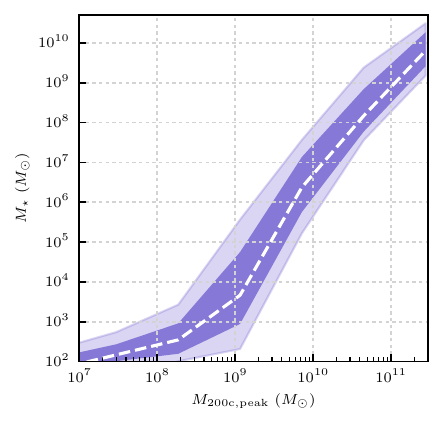}
  \caption[]{Stellar mass of model galaxies as a function of peak virial mass, $M_{\rm 200c,peak}$ of their host halos. The dashed line shows the median stellar mass in bins of halo mass, while the dark and light shaded bands show the 58th and 96th percentiles of $M_\star$ distribution. }
   \label{fig:m200_mstar}
\end{figure}
%

%--------------------------------------------
\subsection{Stellar mass--halo mass relation of model galaxies}
\label{sec:m200_mstar}
%---------------------------------------------

Figure~\ref{fig:m200_mstar} shows the relation between the stellar mass of model galaxies and the peak virial mass of their host halos for the stellar mass range comparable to that of the observed galaxy sample that the model is compared to in this study. The peak mass is estimated as the maximum virial mass along the halo evolution track. The physical origin of this relation and its shape are discussed in \citet{Kravtsov.Manwadkar.2022} and \citet{Manwadkar.Kravtsov.2022} and I refer the interested reader to these papers for details. I note that this relation agrees with existing direct observational constraints on stellar and halo masses of galaxies in this mass range and with the luminosity function of Milky Way satellites, as well as with the relations in the current state-of-the-art cosmological simulations of dwarf galaxy formation \citep{Manwadkar.Kravtsov.2022}.  

Here I reproduce the relation for the model galaxy samples used in this study to illustrate the halo masses in which model galaxies of a given stellar mass reside. In particular, most of the UDGs I will examine have stellar masses of $M_\star\approx 5\times 10^7-10^9\,M_\odot$, which corresponds to halo masses of $M_{\rm 200c}\approx 8\times 10^{9}-2\times 10^{11}\, M_\odot$. This is broadly consistent with the constraints on UDG halo masses of $M_{\rm 200c}\lesssim 10^{11.8}\ M_\odot$ using stacked weak lensing signal \citep{Sifon.etal.2018}, estimates of their halo masses using halo mass-globular cluster relation \citep{Prole.etal.2019}. This is also consistent with the halo mass range $M_{\rm 200c}<10^{11.2}\, M_\odot$ of model UDGs in the Illustris TNG  \citep{Benavides.etal.2023,Doppel.etal.2024}. 

I will also compare these galaxies to dwarf satellite galaxies of the Milky Way, which span a much wider range of masses down to $M_\star\approx 10^3\,M_\odot$ or, according to Figure~\ref{fig:m200_mstar}, halos of masses $M_{\rm 200c}\approx 10^8-10^9\,M_\odot$. Comparisons of the model $M_{\rm tot}(<r_{1/2})$ and $M_{\rm dm}(<r_{1/2})$ with the values estimated for observed galaxies of the same stellar mass will serve as a test of whether model forms galaxies in halos of correct mass and concentration. 

% %
\begin{figure}
  \centering
  \includegraphics[width=0.499\textwidth]{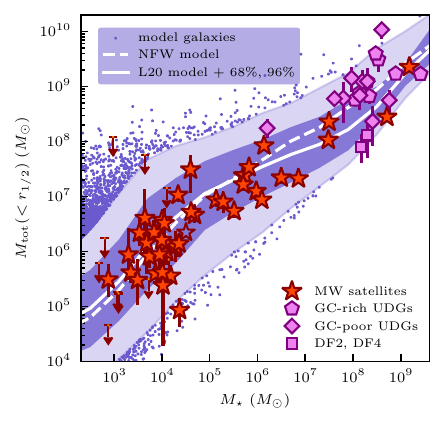}
  \caption[]{Total mass within half-light radius vs stellar mass of model galaxies and dwarf satellites of the Milky Way (stars and upper limits) and UDGs (the sample shown in Figure~\ref{fig:mstar_rhalf}). The dashed and solid lines show the median relations for model galaxies for $M_{\rm tot}(<r_{1/2})$ computed assuming NFW and \citet{Lazar.etal.2020} dark matter density profiles, respectively. The dark- and light-shaded bands show the 68th and 96th percentiles of the distribution around the median of the latter model; blue dots show individual galaxies outside these ranges. The model matches the $M_\star-M_{\rm tot}(<r_{1/2})$ relation well at all stellar masses. The UDGs tend to lie above the median model relation. The DM-deficient galaxies DF2 and DF4 lie below the median relation but within the 96\% of the model galaxy distribution for their stellar mass.} 
   \label{fig:mstar_mhalf}
\end{figure}
%

% %
\begin{figure}
  \centering
  \includegraphics[width=0.499\textwidth]{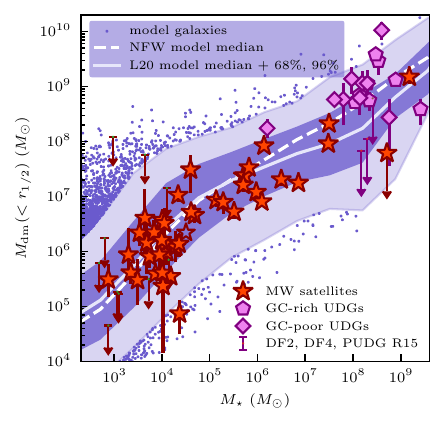}
  \caption[]{Dark matter mass within half-light radius, $M_{\rm dm}(<r_{1/2})=M_{\rm tot}(<r_{1/2})-M_\star/2
  $ vs stellar mass of model galaxies and dwarf satellites of the Milky Way (stars and light red upper limits) and UDGs (squares, diamonds, pentagons, and dark red upper limits). $M_{\rm dm}(<r_{1/2})$ is computed as $M_{\rm tot}(<r_{1/2})-M_\star/2$. The galaxies for which $M_{\rm dm}(<r_{1/2})-\sigma_{M_{\rm dm}}<0$ are shown as upper limits with the top of the limit corresponding to $M_{\rm dm}(<r_{1/2})+\sigma_{M_{\rm dm}}$. The dashed and solid lines show the median relations for model galaxies for $M_{\rm tot}(<r_{1/2})$ computed assuming NFW and \citet{Lazar.etal.2020} dark matter density profiles, respectively. The dark- and light-shaded bands show the 16th and 96th percentiles of the distribution around the median of the latter model; blue dots show individual galaxies outside these ranges. The model matches the $M_\star-M_{\rm dm}(<r_{1/2})$ relation followed by observed galaxies well at all stellar masses. The UDGs tend to lie above the median model relation. The DM-deficient galaxies DF2 and DF4 are shown as upper limits as their $M_{\rm dm}(<r_{1/2})$ is consistent with zero within its uncertainty. This is also true for PUDG-R15 UDG from the sample of \citet{Gannon.etal.2022}, shown as the third dark red upper limit, and for SMC shown as the star with light red upper limit. However, the figure shows that these upper limits are fully consistent with the expected distribution of $M_{\rm dm}(<r_{1/2})$ at the stellar mass of these galaxies. They are thus not DM-deficient in the context of this model. }
   \label{fig:mstar_mhalfdm}
\end{figure}
%

% %
\begin{figure}
  \centering
  \includegraphics[width=0.499\textwidth]{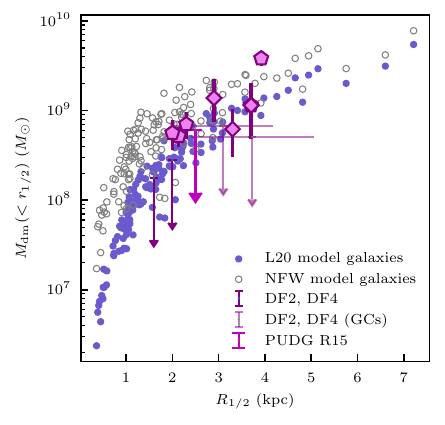}
  \caption[]{Dark matter mass within half-light radius, $M_{\rm dm}(<r_{1/2})$ vs half-light radius for model galaxies in the stellar mass range of $7\times 10^7<M_\star/M_\odot<3\times 10^8\, M_\odot$ assuming NFW (open gray circles) and \citet[][blue circles]{Lazar.etal.2020} dark matter density profiles. The 95\% upper limits shown with darker arrows represent constraints on  $M_{\rm dm}(<r_{1/2})$ for observed galaxies DF2, DF4 and PUDG-R15. For comparison lighter arrows show mass constraints from GC velocity dispersion at the GC half-number radii in DF2 and DF4 galaxies; the horizontal bars indicate uncertainty in their half-number radii. Other UDGs in this mass range are shown as points with the same type as in Figures~\ref{fig:mstar_mhalf} and \ref{fig:mstar_mhalfdm}. The galaxy with the largest $R_{1/2}$ and $M_{\rm dm}$ is DF44.}
   \label{fig:mdm_rhalf}
\end{figure}

\begin{figure}
  \centering
  \includegraphics[width=0.499\textwidth]{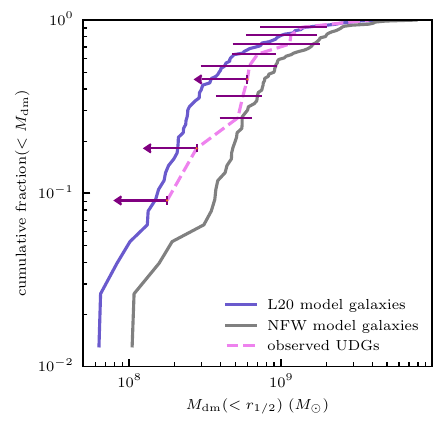}
  \caption[]{Cumulative distribution of dark matter mass within half-light radius, $M_{\rm dm}(<r_{1/2})$ for galaxies in the stellar mass range of $7\times 10^7<M_\star/M_\odot<3\times 10^8\, M_\odot$. The model galaxies for the NFW halo mass profile model are shown by the gray line and the \citet[][]{Lazar.etal.2020} model is shown by the blue line. The dashed magenta line shows the cumulative distribution for $M_{\rm dm}(<r_{1/2})$ for 10 UDG galaxies with measured stellar velocity dispersions in this stellar mass range. The cumulative distribution in this case is constructed by sorting galaxies in their formal estimated $M_{\rm dm}(<r_{1/2})$ or its 95\% upper limit. The horizontal bars indicate the uncertainty of each $M_{\rm dm}(<r_{1/2})$ estimate while upper limits (for DF2, DF4 and PUDG-R15) are shown by leftward arrows. }
   \label{fig:mdm_dist}
\end{figure}

%-------------------
\section{Results}
\label{sec:results}
%-------------------

Figure~\ref{fig:mstar_mhalf} shows the total mass within half-light radius for model galaxies and observational estimates for the MW dwarf satellites, DM-deficient UDGs DF2 and DF4, and the sample of UDGs with measured stellar velocity dispersions compiled by \citet{Gannon.etal.2024cat}. Note that I chose to convert observational estimates of velocity dispersion to $M_{\rm tot}(<r_{1/2})$ and $M_{\rm dm}(<r_{1/2})$ instead of attempting to model $\sigma_{\star,\rm los}$ because such modeling would require additional assumptions. In the context of the semi-analytic galaxy formation model used here, these assumptions  would be equally or more uncertain than uncertainties associated with using the conversion of the observational estimates of $\sigma_{\star,\rm los}$ and $R_{1/2}$ to the total and dark matter masses within $r_{1/2}$.

Figure~\ref{fig:mstar_mhalf} shows a remarkable agreement between the $\Lambda$CDM-based galaxy formation model results (lines and shaded bands) and the relation traced by the observed galaxies shown as points. 
The figure also shows that UDGs follow the relation of the Milky Way dwarf satellites. 

The UDGs are also broadly consistent with model expectations, but I note that quite a few are located outside the $1\sigma$ model band.  On the high side, DGSAT-1 UDG is outside $2\sigma$ band and DF44 is close to it. The Small Magellanic Cloud (SMC) galaxy has comparable baryon and dark matter masses within a half-light radius (see below), while the Nube galaxy with a similar stellar mass and half-light radius of 6.9 kpc is dark matter dominated \citep{Montes.etal.2024}.

One can note that $M_{\rm tot}(<r_{1/2})$ becomes close to $M_\star$ for galaxies with $M_\star>10^8\, M_\odot$. This is due to a general trend of decreasing $M_{\rm tot}(<r_{1/2})$-to-luminosity ratio with increasing luminosity for dwarf galaxies \citep[e.g.,][]{Wolf.etal.2010}. Specifically, the model predicts that galaxies of $M_\star\lesssim 3\times 10^8\,M_\odot$ are dark matter dominated ($M_{\rm dm}(<r_{1/2})>M_\star/2$), while in larger mass galaxies mass within $r_{1/2}$ is dominated by stars. Gas potentially can shift the transition between these regimes to somewhat smaller $M_\star$.

Nevertheless, each model galaxy has dark matter within $r_{1/2}$ and thus $M_{\rm tot}(<r_{1/2})>M_\star/2$. Figure~\ref{fig:mstar_mhalfdm} shows $M_{\rm dm}(<r_{1/2})=M_{\rm tot}(<r_{1/2})-M_\star/2$ as a function of $M_\star$. In most observed galaxies $M_{\rm dm}(<r_{1/2})-\sigma_{M_{\rm dm}}>0$ and their $M_{\rm dm}(<r_{1/2})$ are shown as points with errobars. 
Galaxies with $M_{\rm dm}(<r_{1/2})-\sigma_{M_{\rm dm}}<0$ are shown as upper limits with the top of the limit corresponding to $M_{\rm dm}(<r_{1/2})+\sigma_{M_{\rm dm}}$. 

The figure shows that there are only 4 such galaxies: three UDGs -- DF2, DF4 \citep[``dark matter-deficient'' UDGs in the NGC 1052 group,][]{Danieli.etal.2019,Shen.etal.2023} and PUDG-R15 \citep[a UDG in the Perseus cluster][]{Gannon.etal.2022} -- and SMC.\footnote{The SMC has the half-light radius of $1106\pm 77$ pc \citep{Munoz.etal.2018} and stellar mass within this radius of $1.5-2.5\times 10^8\, M_\odot$ \citep[given the estimates of total stellar mass of $3-5\times 10^8\, M_\odot$][]{Harris.Zaritsky.2004,Skibba.etal.2012,DiTeodoro.etal.2019}, while its rotation velocity of $33\pm 2$ km/s \citep{DiTeodoro.etal.2019} at the half-light radius implies $M_{\rm tot}(<r_{1/2})=2.8\pm 0.4\times 10^8\,M_\odot$.} In all four of these galaxies $M_{\rm dm}(<r_{1/2})$ is consistent with zero. However, Figure~\ref{fig:mstar_mhalfdm} shows that the upper limits on $M_{\rm dm}(<r_{1/2})$ in these galaxies are also consistent with expectation of the $\Lambda$CDM-based galaxy formation model. In other words, the amount of the expected DM within half-light radii of these galaxies is consistent with observational upper limits. 

Galaxies of a given stellar mass have a broad range of sizes which translates to a broad range of $M_{\rm tot}(<r_{1/2})$ at a given stellar mass \citep[as also noted for observed UDGs by][]{vanDokkum.etal.2019b}. Given that $M_{\rm tot}(<r_{1/2})$ depends on $r_{1/2}$, one may wonder if the model galaxies that have $M_{\rm dm}(<r_{1/2})$ consistent with the constraints of DF2, DF4, and PUDG-R15 have predominantly sizes smaller than the sizes of these galaxies. 

Figure~\ref{fig:mdm_rhalf} shows that this is not the case. This figure compares $M_{\rm dm}(<r_{1/2})$ as a function of $R_{1/2}$ for observed and model galaxies in the stellar mass range of $7\times 10^7<M_\star/M_\odot<3\times 10^8\, M_\odot$. The dark matter mass for model galaxies is shown for both the NFW and \citet[][]{Lazar.etal.2020} dark matter density profiles. Figure~\ref{fig:mdm_rhalf} shows that the upper limits for DF2, DF4, and PUDG-R15 galaxies are consistent with $M_{\rm dm}(<r_{1/2})$ in model galaxies of comparable size, even though the upper limits on dark matter mass for these observed galaxies were estimated without taking into account uncertainty of their stellar mass estimates. This is also true for constraints from GC velocity dispersion at the GC half-number radii in DF2 and DF4 galaxies shown by lighter arrows. Note that the horizontal bars indicate uncertainty in the half-number radii in these galaxies, estimated using bootstrap resampling of the projected radii of 10 and 7 GCs in these two galaxies, respectively.
Other UDGs are also consistent with model expectations and lie on the $M_{\rm dm}-R_{1/2}$ relation traced by model galaxies, except DF44 galaxy that lies considerably larger than relation for the \citet[][]{Lazar.etal.2020} model, although it is still close to the mass range expected for the NFW model. This may indicate that feedback effects on the mass profile were particularly weak in this galaxy. 

The overall dependence of $M_{\rm dm}(<r_{1/2})$ on $r_{1/2}$ reflects the dark matter mass profile of halos hosting galaxies of this stellar mass.  The effect of feedback on the mass profile at this stellar mass ($M_\star\approx 10^8\, M_\odot$) quantified by the \citet{Lazar.etal.2020} profile is larger than the scatter of $M_{\rm dm}$ at a given $R_{1/2}$. This effect brings model masses in better agreement with observational estimates of $M_{\rm tot}(<r_{1/2})$ for UDGs (except for DF44), although even unmodified NFW mass profiles are still fairly consistent with these estimates. 

Figure~\ref{fig:mdm_rhalf} shows that the $M_{\rm dm}(<r_{1/2})$ of UDGs span a narrow range of values than model galaxies of the same stellar mass. This is because these galaxies are selected to have large sizes, $R_{1/2}> 1.5$ kpc, and thus enclose larger mass. Figure~\ref{fig:mdm_dist} shows cumulative distributions of $M_{\rm dm}(<r_{1/2})$ for the model galaxies in the same stellar mass range, but subject to the same size cut (gray and blue solid lines for the NFW and \citealt{Lazar.etal.2020} model, respectively). The dashed magenta line shows the cumulative distribution for $M_{\rm dm}(<r_{1/2})$ for 10 UDG galaxies with measured stellar velocity dispersions in this stellar mass range. The cumulative distribution in this case is constructed by sorting galaxies in their formal estimated $M_{\rm dm}(<r_{1/2})$ or its 95\% upper limit. The horizontal bars indicate the uncertainty of each $M_{\rm dm}(<r_{1/2})$ estimate while upper limits (for DF2, DF4 and PUDG-R15) are shown by leftward arrows. Figure~\ref{fig:mdm_rhalf} shows that observed distribution for observed UDG galaxies is broadly consistent with the model, although the number of galaxies is small and a similar tests should be done when larger samples with dynamical mass estimates become available. 

The scatter of $M_{\rm dm}$ for a given value of $R_{1/2}$ for the \citet{Lazar.etal.2020} mass profile is due to 1) scatter of halo concentrations and 2) scatter of $M_\star/M_{\rm 200c}$ due to the dependence of the \citet{Lazar.etal.2020} profile on this ratio. For the galaxies of $M_\star\sim 10^8\,M_\odot$ with sizes of $1.5<r_{1/2}<3$ kpc shown in Figure~\ref{fig:mdm_rhalf}, there is a positive correlation between $M_{\rm dm}(<r_{1/2})$ and concentration $c_{\rm 200c}$ with the correlation coefficient of $0.51\pm 0.09$, where uncertainty is estimated using bootstrap resampling. 
%Furthermore, for galaxies of a given stellar mass, the scatter in the $M_\star-M_{\rm 200c}$ relation results in a range of halo masses and size in the galaxy formation model used here correlates with halo virial radius and mass. For example, for the galaxies shown in Figure \ref{fig:mdm_rhalf}, there is a correlation of $r_{1/2}$ and $R_{\rm 200c}$ with the correlation coefficient of $0.31\pm 0.08$. 
There is also a weaker anti-correlation between $M_{\rm dm}(<r_{1/2})$ and $\log_{10}M_\star/M_{\rm 200c}$ with the correlation coefficient of $-0.31\pm 0.11$. 

Thus, in the galaxy formation model used here, the most dark matter-deficient galaxies of a given size correspond to halos with the smallest concentrations and the largest ratios of $M_\star/M_{\rm 200c}$. Conversely, the most dark matter-dominated galaxies are hosted by the highest concentration halos with the smallest $M_\star/M_{\rm 200c}$ ratios. Note that halo concentration correlates with halo formation time with low (high) concentration halos forming late (early). The $M_\star/M_{\rm 200c}$ ratio depends on the interplay between gas accretion, star formation, and feedback-driven outflows for different halo mass assembly histories. 

%-------------------
\section{Discussion}
\label{sec:discussion}
%-------------------

The main conclusion that can be drawn from the results presented above is that the dark matter content of UDGs  is broadly consistent with the expectation of our $\Lambda$CDM-based galaxy formation model. The large variation in the amount of dark matter contribution to the total mass within $r_{1/2}$ for observed dwarf galaxies and UDGs of the same stellar mass -- from very dark matter-dominated galaxies DGSAT-1 and DF44 to ``dark matter-deficient'' galaxies DF2, DF4, and PUDG-R15 -- is due mainly to 1) the large range of sizes of galaxies of a given $M_\star$ (see Figure~\ref{fig:mstar_rhalf}), 2) the concentration of their halo and 3) feedback effects expected for the galaxies of such stellar mass \citep[e.g.,][]{Governato.etal.2012,Pontzen.Governato.2012,DiCintio.etal.2014,Chan.etal.2015,Tollet.etal.2016,Read.etal.2016,Lazar.etal.2020,Brook.etal.2021}. This variation is thus much larger than the expected scatter in the dark matter density profiles at a {\it fixed} radius for galaxies of a given stellar mass \citep[as is the case in observed UDGs][]{vanDokkum.etal.2019b}. 

\begin{figure}
  \centering
  \includegraphics[width=0.499\textwidth]{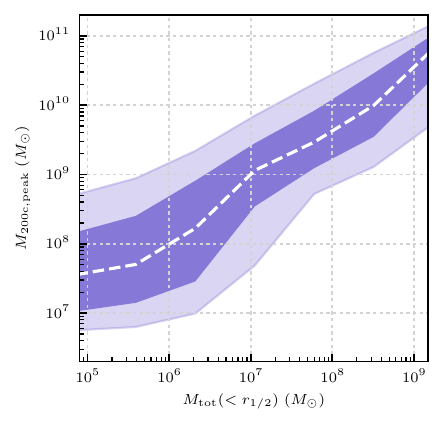}
  \caption[]{Relation between total mass within half-light radius, $M_{\rm tot}(<r_{1/2})$, estimated using the \citet{Lazar.etal.2020} dark matter mass profile model, and the peak virial mass of the halo, $M_{\rm 200c,peak}$. The dashed line shows the median of the distribution of $M_{\rm 200c,peak}$ in bins of $M_{\rm tot}(<r_{1/2})$, while dark- and light-shaded bands show 68th and 96th percentiles of its distribution. The figure show that a well-defined relation between $M_{\rm tot}(<r_{1/2})$ and $M_{\rm 200c,peak}$ exists for model galaxies, but the scatter around the median relation is significant.}
   \label{fig:mhalf_m200}
\end{figure}

For these reasons, the range of virial halo masses for a given $M_{\rm tot}(<r_{1/2})$ is also quite broad. Figure~\ref{fig:mhalf_m200} shows the peak virial mass, $M_{\rm 200c,peak}$ of host halos hosting model galaxies for galaxies with a given $M_{\rm tot}(<r_{1/2})$ estimated using the \citet{Lazar.etal.2020} dark matter mass profile model. The median of the distribution is shown by the dashed line, while $68$th and $96$th percentiles are shown by the dark and light shaded areas. Although there is a clear relationship between the two masses, the $96$th percentile range of the virial mass spans $\gtrsim 1.5$ dex at a given $M_{\rm tot}(<r_{1/2})$. Specifically, galaxies with $M_{\rm tot}(<r_{1/2})=5\times 10^8\,M_\odot$ typical for the UDGs shown in the figures can be hosted in halos of $M_{\rm 200c}$ as low as $\approx 3\times 10^9\,M_\odot$ and as high as $\approx 8\times 10^{10}\, M_\odot$. Still, the halo mass range of observed UDGs indicated by our model implies that these galaxies are hosted by halos of sub-LMC mass, consistent with conclusions of \citet{Rong.etal.2017} based on the analyses of the Millenium simulation coupled with a semi-analytic galaxy formation model. This is also consistent with the lack of X-rays from hot gas and X-ray binaries from UDGs that were believed to be most massive \citep{Kovacs.etal.2019,Bogdan.2020}.

% %

The origin of UDGs, and especially the origin of the dark matter-deficient UDGs, is still actively debated. One class of models for abundant UDGs in cluster environment considers ram pressure stripping of gas, accompanied by subsequent suppression of gas accretion by the intracluster medium and tidal heating \citep{Yozin.Bekki.2015,Safarzadeh.Scannapieco.2017,Carleton.etal.2019,Tremmel.etal.2020,Sales.etal.2020,Moreno.etal.2022,Fielder.etal.2024}. Another class invokes internal processes, such as  inefficient cooling \citep{Wang.etal.2023ghostly}, leading to large sizes and low surface brightness and star formation. \citet{Wright.etal.2021} find that in their simulations field UDGs form via early mergers, which induce high spin onto the gas and temporarily boost and redistribute gas and star formation to the outskirts of galaxies, resulting in lower central SFRs and lower surface brightness later on. 
Finally, remnants of high-speed collisions of galaxies were proposed as a possible origin for dark-matter deficient UDGs  \citep{Baushev.2018} and this scenario is supported by simulation results \citep{Lee.etal.2021,Lee.etal.2024}.  \citet{vanNest.etal.2022} point out that the interpretation of the UDG formation may depend on the details of the UDG definition and may be different for diffuse galaxies in clusters and in the field. Furthermore, a combination of various processes may be at play during UDG formation \citep[e.g.,][]{Jackson.etal.2021}.

\citet{vanDokkum.etal.2022} argued that DF2 and DF4 UDGs may have formed from gas stripped during a collision of a dwarf galaxy with a massive galaxy NGC 1052. The common formation process of these galaxies is indeed indicated by the similarity of colors of their globular clusters \citep{vanDokkum.etal.2022b}.
The tidal dwarf formation scenario for dark matter-deficient UDGs was also explored by \citet{Ivleva.etal.2024} but in their scenario UDGs form from the gas stripped from galaxies by the ram pressure from the intracluster medium. \citet[][]{Keim.etal.2022} present evidence that DF2 and DF4 UDGs have tidal features and thus may be undergoing a significant tidal stripping, which may well have reduced the dark matter mass within their half-light radius. \citet{Golini.etal.2024}, however, while confirming the presence of tidal features in DF4, find no indications of tidal distorsions in DF2. 
  
The results presented in this paper do not exclude any of the specific formation scenarios discussed in the literature. In particular, they do not change the fact that DF2 and DF4 galaxies share several similarities, including the abundance and colors of their GCs. Results presented here only imply that no special formation mechanism is required to explain the DM content of UDGs because their current dark matter content constraints are consistent with the range of dark matter masses expected for halos hosting UDGs, when the expected effect of stellar feedback on the dark matter density profiles is taken into account. In addition to the effects on dark matter profile, this feedback can also heat stellar population of galaxies and help explain the large sizes and low surface brightnesses of UDGs \citep{DiCintio.etal.2017,Chan.etal.2018,Martin.etal.2019,Jiang.etal.2019,Jackson.etal.2021,Trujillo_Gomez.etal.2021,Trujillo_Gomez.etal.2022}. 

Overall, models of galaxies do expect a galaxy population with a wide range of sizes 
\citep[][cf. also \citealt{Dalcanton.etal.1997}, \citealt{Rong.etal.2017,Rong.etal.2024}, \citealt{Liao.etal.2019} and \citealt{Benavides.etal.2023}]{Amorisco.Loeb.2016}. Observed galaxies of a given stellar mass also span a broad range of sizes (see, e.g., Fig 1 above). Recent observational studies find that UDGs in clusters correspond {\it structurally} to the low-surface brightness tail of the regular dwarf galaxy population \citep{Martin.etal.2018,Trujillo.2021,Zoeller.etal.2024,Marleau.etal.2024}. Nevertheless, many UDGs also have unusually large GC populations with unusual luminosity functions despite their low surface densities. This indicates that these galaxies' evolution differs from that of a typical low-surface brightness galaxy in some crucial aspects. 

Note that the model used here also includes feedback and related gas outflows and star formation regulation that have been invoked as a mechanism explaining the formation of UDGs, as well as strangulation whereby galaxies in halos that stop growing mass also stop accreting fresh gas. 

As a final remark, the model presented in this study does not address the interesting issue of globular cluster populations of UDGs. Given that many UDGs (and, in particular, DM-deficient UDGs) contain anomalously large GC populations, often with unusual luminosity functions, a comprehensive model for their formation must account for the diversity of GC populations in these galaxies. Such models would be instrumental about understanding of the physical origin of both large sizes and anomalous GC populations in these fascinating systems.    

%---------------------
\section{Conclusions}
\label{sec:conclusions}
%---------------------

I have compared the dark matter content within a half-mass radius in model galaxies formed within a realistic $\Lambda$CDM-based galaxy formation model and observed dwarf satellites of the Milky Way and ultra-diffuse galaxies. The model reproduces the main properties and scaling relations of dwarf galaxies, in particular their stellar mass-size relation \citep[][see also Fig.~\ref{fig:mstar_rhalf} above]{Manwadkar.Kravtsov.2022}. The main results and conclusions of this study can be summarized as follows. 

\begin{itemize}
    \item[1.] The model reproduces the total mass, $M_{\rm tot}(<r_{1/2})=M_\star/2+M_{\rm dm}(<r_{1/2})$, and dark matter mass, $M_{\rm dm}(<r_{1/2})$, within half-mass radius of observed dwarf galaxies. It also reproduces well the correlation of these masses with stellar mass exhibited by observed galaxies (see Figures~\ref{fig:mstar_mhalf} and \ref{fig:mstar_mhalfdm}).
    \item[2.] Ultra-diffuse galaxies lie on the same $M_{\rm dm}(<r_{1/2})-M_\star$ relation as the Milky Way dwarf satellites but their masses are generally larger than the median expected mass due to their large sizes (Figures~\ref{fig:mstar_mhalfdm} and \ref{fig:mdm_rhalf}). 
    \item[3.] The scatter in the $M_{\rm dm}(<r_{1/2})-M_\star$ relation is driven primarily by the broad range of sizes of galaxies of a given stellar mass (Figure \ref{fig:mdm_rhalf}). This large scatter results in a wide range of halo virial masses that may correspond to a given value of mass within $r_{1/2}$ (Figure \ref{fig:mhalf_m200}). 
    \item[4.] The most dark matter-deficient galaxies of a given size correspond to halos with the smallest concentrations and the largest ratios of $M_\star/M_{\rm 200c}$. Conversely, the most dark matter-dominated galaxies are hosted by the highest concentration halos with the smallest $M_\star/M_{\rm 200c}$ ratios.
    \item[5.]  The model indicates that UDGs with stellar mass of $M_\star\approx 10^8\,M_\odot$ form in halos of the present-day virial mass $M_{\rm 200c}\approx 10^{10}-10^{11}\,M_\odot$ (see Figure \ref{fig:m200_mstar}). The scatter between $M_{\rm dm}(<r_{1/2})$  and $M_{\rm 200c}$, however, is expected to be large (Figure \ref{fig:mhalf_m200}, which renders inference of the virial mass from  $M_{\rm dm}(<r_{1/2})$ uncertain and dependent on specific assumptions about the halo mass profile.
\end{itemize}

Results presented in this paper indicate that dark matter-deficient UDGs may represent a tail of the expected dark matter profiles, especially if the effect of feedback on these profiles is taken into account. Nevertheless, the unusually rich GC populations in some UDG galaxies do indicate unusual evolution. Our results simply indicate that with the current accuracy of the velocity dispersion measurements their $M_{\rm dm}(<r_{1/2})$ are consistent with the range expected in the $\Lambda$CDM model. It will be important to test this conclusion in the future using larger samples of ultra-diffuse galaxies and more accurate constraints on the masses with half-mass radii. 

\section*{Acknowledgements}
I am very grateful to Shany Danieli for her incisive comments on the draft of this paper and to Duncan Forbes for comments that allowed clarification of some of the results and interpretations. I would also like to thank two referees -- Pieter van Dokkum and an anonymous referee -- for constructive comments that improved the paper.
I thank Alexander Ji and the Caterpillar collaboration for providing halo tracks of the Caterpillar simulations used in this study and  Shea Garrison-Kimmel and Michael Boylan-Kolchin for providing halo tracks of the ELVIS and Phat ELVIS simulations.
AK was supported by the National Science Foundation grant AST-1911111 and NASA ATP grant 80NSSC20K0512. 
Analyses presented in this paper were greatly aided by the following free software packages: {\tt NumPy} \citep{NumPy2015,numpy}, {\tt SciPy} \citep{SciPy}, and {\tt Matplotlib} \citep{matplotlib}. I have also used the Astrophysics Data Service (\href{http://adsabs.harvard.edu/abstract_service.html}{\tt ADS}) and \href{https://arxiv.org}{\tt arXiv} preprint repository extensively during this project and the writing of the paper.

\section*{Data Availability}
Halo catalogs from the Caterpillar simulations are available at \href{https://www.caterpillarproject.org/}{\tt https://www.caterpillarproject.org/}. The \texttt{GRUMPY} model pipeline is available at \url{https://github.com/kibokov/GRUMPY}. The data used in the plots within this article are available on request to the author.

\bibliographystyle{mnras}

\bibliography{mhalf_udg}

\begin{thebibliography}{}
\makeatletter
\relax
\def\mn@urlcharsother{\let\do\@makeother \do\$\do\&\do\#\do\^\do\_\do\%\do\~}
\def\mn@doi{\begingroup\mn@urlcharsother \@ifnextchar [ {\mn@doi@}
  {\mn@doi@[]}}
\def\mn@doi@[#1]#2{\def\@tempa{#1}\ifx\@tempa\@empty \href
  {http://dx.doi.org/#2} {doi:#2}\else \href {http://dx.doi.org/#2} {#1}\fi
  \endgroup}
\def\mn@eprint#1#2{\mn@eprint@#1:#2::\@nil}
\def\mn@eprint@arXiv#1{\href {http://arxiv.org/abs/#1} {{\tt arXiv:#1}}}
\def\mn@eprint@dblp#1{\href {http://dblp.uni-trier.de/rec/bibtex/#1.xml}
  {dblp:#1}}
\def\mn@eprint@#1:#2:#3:#4\@nil{\def\@tempa {#1}\def\@tempb {#2}\def\@tempc
  {#3}\ifx \@tempc \@empty \let \@tempc \@tempb \let \@tempb \@tempa \fi \ifx
  \@tempb \@empty \def\@tempb {arXiv}\fi \@ifundefined
  {mn@eprint@\@tempb}{\@tempb:\@tempc}{\expandafter \expandafter \csname
  mn@eprint@\@tempb\endcsname \expandafter{\@tempc}}}

\bibitem[\protect\citeauthoryear{{Alabi} et~al.,}{{Alabi}
  et~al.}{2018}]{Alabi2018}
{Alabi} A.,  et~al., 2018, \mn@doi [MNRAS] {10.1093/mnras/sty1616}, \href
  {https://ui.adsabs.harvard.edu/abs/2018MNRAS.479.3308A} {479, 3308}

\bibitem[\protect\citeauthoryear{{Amorisco} \& {Loeb}}{{Amorisco} \&
  {Loeb}}{2016}]{Amorisco.Loeb.2016}
{Amorisco} N.~C.,  {Loeb} A.,  2016, \mn@doi [\mnras] {10.1093/mnrasl/slw055},
  \href {https://ui.adsabs.harvard.edu/abs/2016MNRAS.459L..51A} {459, L51}

\bibitem[\protect\citeauthoryear{{Baushev}}{{Baushev}}{2018}]{Baushev.2018}
{Baushev} A.~N.,  2018, \mn@doi [\na] {10.1016/j.newast.2017.10.008}, \href
  {https://ui.adsabs.harvard.edu/abs/2018NewA...60...69B} {60, 69}

\bibitem[\protect\citeauthoryear{{Bautista}, {Koda}, {Yagi}, {Komiyama}  \&
  {Yamanoi}}{{Bautista} et~al.}{2023}]{Bautista.etal.2023}
{Bautista} J. M.~G.,  {Koda} J.,  {Yagi} M.,  {Komiyama} Y.,   {Yamanoi} H.,
  2023, \mn@doi [\apjs] {10.3847/1538-4365/acd3e7}, \href
  {https://ui.adsabs.harvard.edu/abs/2023ApJS..267...10B} {267, 10}

\bibitem[\protect\citeauthoryear{{Beasley}, {Romanowsky}, {Pota}, {Navarro},
  {Martinez Delgado}, {Neyer}  \& {Deich}}{{Beasley}
  et~al.}{2016}]{Beasley2016}
{Beasley} M.~A.,  {Romanowsky} A.~J.,  {Pota} V.,  {Navarro} I.~M.,  {Martinez
  Delgado} D.,  {Neyer} F.,   {Deich} A.~L.,  2016, \mn@doi [ApJL]
  {10.3847/2041-8205/819/2/L20}, \href
  {https://ui.adsabs.harvard.edu/abs/2016ApJ...819L..20B} {819, L20}

\bibitem[\protect\citeauthoryear{{Behroozi}, {Wechsler}, {Wu}, {Busha},
  {Klypin}  \& {Primack}}{{Behroozi} et~al.}{2013}]{Behroozi.etal.2013}
{Behroozi} P.~S.,  {Wechsler} R.~H.,  {Wu} H.-Y.,  {Busha} M.~T.,  {Klypin}
  A.~A.,   {Primack} J.~R.,  2013, \mn@doi [\apj] {10.1088/0004-637X/763/1/18},
  \href {https://ui.adsabs.harvard.edu/abs/2013ApJ...763...18B} {763, 18}

\bibitem[\protect\citeauthoryear{{Behroozi}, {Hearin}  \& {Moster}}{{Behroozi}
  et~al.}{2022}]{Behroozi.etal.2022}
{Behroozi} P.,  {Hearin} A.,   {Moster} B.~P.,  2022, \mn@doi [\mnras]
  {10.1093/mnras/stab3193}, \href
  {https://ui.adsabs.harvard.edu/abs/2022MNRAS.509.2800B} {509, 2800}

\bibitem[\protect\citeauthoryear{{Benavides}, {Sales}, {Abadi}, {Marinacci},
  {Vogelsberger}  \& {Hernquist}}{{Benavides}
  et~al.}{2023}]{Benavides.etal.2023}
{Benavides} J.~A.,  {Sales} L.~V.,  {Abadi} M.~G.,  {Marinacci} F.,
  {Vogelsberger} M.,   {Hernquist} L.,  2023, \mn@doi [\mnras]
  {10.1093/mnras/stad1053}, \href
  {https://ui.adsabs.harvard.edu/abs/2023MNRAS.522.1033B} {522, 1033}

\bibitem[\protect\citeauthoryear{{Bogd{\'a}n}}{{Bogd{\'a}n}}{2020}]{Bogdan.2020}
{Bogd{\'a}n} {\'A}.,  2020, \mn@doi [\apjl] {10.3847/2041-8213/abb886}, \href
  {https://ui.adsabs.harvard.edu/abs/2020ApJ...901L..30B} {901, L30}

\bibitem[\protect\citeauthoryear{{Bothun}, {Beers}, {Mould}  \&
  {Huchra}}{{Bothun} et~al.}{1985}]{Bothun.etal.1985}
{Bothun} G.~D.,  {Beers} T.~C.,  {Mould} J.~R.,   {Huchra} J.~P.,  1985,
  \mn@doi [\aj] {10.1086/113951}, \href
  {https://ui.adsabs.harvard.edu/abs/1985AJ.....90.2487B} {90, 2487}

\bibitem[\protect\citeauthoryear{{Bothun}, {Impey}, {Malin}  \&
  {Mould}}{{Bothun} et~al.}{1987}]{Bothun.etal.1987}
{Bothun} G.~D.,  {Impey} C.~D.,  {Malin} D.~F.,   {Mould} J.~R.,  1987, \mn@doi
  [\aj] {10.1086/114443}, \href
  {https://ui.adsabs.harvard.edu/abs/1987AJ.....94...23B} {94, 23}

\bibitem[\protect\citeauthoryear{{Bothun}, {Impey}  \& {McGaugh}}{{Bothun}
  et~al.}{1997}]{Bothun.etal.1997}
{Bothun} G.,  {Impey} C.,   {McGaugh} S.,  1997, \mn@doi [\pasp]
  {10.1086/133941}, \href
  {https://ui.adsabs.harvard.edu/abs/1997PASP..109..745B} {109, 745}

\bibitem[\protect\citeauthoryear{{Brook}, {Di Cintio}, {Macci{\`o}}  \&
  {Blank}}{{Brook} et~al.}{2021}]{Brook.etal.2021}
{Brook} C.~B.,  {Di Cintio} A.,  {Macci{\`o}} A.~V.,   {Blank} M.,  2021,
  \mn@doi [\apjl] {10.3847/2041-8213/ac236a}, \href
  {https://ui.adsabs.harvard.edu/abs/2021ApJ...919L...1B} {919, L1}

\bibitem[\protect\citeauthoryear{{Carleton}, {Errani}, {Cooper}, {Kaplinghat},
  {Pe{\~n}arrubia}  \& {Guo}}{{Carleton} et~al.}{2019}]{Carleton.etal.2019}
{Carleton} T.,  {Errani} R.,  {Cooper} M.,  {Kaplinghat} M.,  {Pe{\~n}arrubia}
  J.,   {Guo} Y.,  2019, \mn@doi [\mnras] {10.1093/mnras/stz383}, \href
  {https://ui.adsabs.harvard.edu/abs/2019MNRAS.485..382C} {485, 382}

\bibitem[\protect\citeauthoryear{{Carlsten}, {Greene}, {Beaton}, {Danieli}  \&
  {Greco}}{{Carlsten} et~al.}{2022}]{Carlsten.etal.2022}
{Carlsten} S.~G.,  {Greene} J.~E.,  {Beaton} R.~L.,  {Danieli} S.,   {Greco}
  J.~P.,  2022, \mn@doi [\apj] {10.3847/1538-4357/ac6fd7}, \href
  {https://ui.adsabs.harvard.edu/abs/2022ApJ...933...47C} {933, 47}

\bibitem[\protect\citeauthoryear{{Chan}, {Kere{\v{s}}}, {O{\~n}orbe},
  {Hopkins}, {Muratov}, {Faucher-Gigu{\`e}re}  \& {Quataert}}{{Chan}
  et~al.}{2015}]{Chan.etal.2015}
{Chan} T.~K.,  {Kere{\v{s}}} D.,  {O{\~n}orbe} J.,  {Hopkins} P.~F.,  {Muratov}
  A.~L.,  {Faucher-Gigu{\`e}re} C.~A.,   {Quataert} E.,  2015, \mn@doi [\mnras]
  {10.1093/mnras/stv2165}, \href
  {https://ui.adsabs.harvard.edu/abs/2015MNRAS.454.2981C} {454, 2981}

\bibitem[\protect\citeauthoryear{{Chan}, {Kere{\v{s}}}, {Wetzel}, {Hopkins},
  {Faucher-Gigu{\`e}re}, {El-Badry}, {Garrison-Kimmel}  \&
  {Boylan-Kolchin}}{{Chan} et~al.}{2018}]{Chan.etal.2018}
{Chan} T.~K.,  {Kere{\v{s}}} D.,  {Wetzel} A.,  {Hopkins} P.~F.,
  {Faucher-Gigu{\`e}re} C.~A.,  {El-Badry} K.,  {Garrison-Kimmel} S.,
  {Boylan-Kolchin} M.,  2018, \mn@doi [\mnras] {10.1093/mnras/sty1153}, \href
  {https://ui.adsabs.harvard.edu/abs/2018MNRAS.478..906C} {478, 906}

\bibitem[\protect\citeauthoryear{{Chilingarian}, {Afanasiev}, {Grishin},
  {Fabricant}  \& {Moran}}{{Chilingarian} et~al.}{2019}]{Chilingarian2019}
{Chilingarian} I.~V.,  {Afanasiev} A.~V.,  {Grishin} K.~A.,  {Fabricant} D.,
  {Moran} S.,  2019, \mn@doi [ApJ] {10.3847/1538-4357/ab4205}, \href
  {https://ui.adsabs.harvard.edu/abs/2019ApJ...884...79C} {884, 79}

\bibitem[\protect\citeauthoryear{{Collins}, {Tollerud}, {Rich}, {Ibata},
  {Martin}, {Chapman}, {Gilbert}  \& {Preston}}{{Collins}
  et~al.}{2020}]{Collins.etal.2020}
{Collins} M. L.~M.,  {Tollerud} E.~J.,  {Rich} R.~M.,  {Ibata} R.~A.,  {Martin}
  N.~F.,  {Chapman} S.~C.,  {Gilbert} K.~M.,   {Preston} J.,  2020, \mn@doi
  [MNRAS] {10.1093/mnras/stz3252}, \href
  {https://ui.adsabs.harvard.edu/abs/2020MNRAS.491.3496C} {491, 3496}

\bibitem[\protect\citeauthoryear{{Conselice}, {Gallagher}  \&
  {Wyse}}{{Conselice} et~al.}{2003}]{Conselice.etal.2003}
{Conselice} C.~J.,  {Gallagher} John~S. I.,   {Wyse} R. F.~G.,  2003, \mn@doi
  [\aj] {10.1086/345385}, \href
  {https://ui.adsabs.harvard.edu/abs/2003AJ....125...66C} {125, 66}

\bibitem[\protect\citeauthoryear{{Dalcanton}, {Spergel}, {Gunn}, {Schmidt}  \&
  {Schneider}}{{Dalcanton} et~al.}{1997}]{Dalcanton.etal.1997}
{Dalcanton} J.~J.,  {Spergel} D.~N.,  {Gunn} J.~E.,  {Schmidt} M.,
  {Schneider} D.~P.,  1997, \mn@doi [\aj] {10.1086/118499}, \href
  {https://ui.adsabs.harvard.edu/abs/1997AJ....114..635D} {114, 635}

\bibitem[\protect\citeauthoryear{{Danieli} \& {van Dokkum}}{{Danieli} \& {van
  Dokkum}}{2019}]{Danieli.vanDokkum.2019}
{Danieli} S.,  {van Dokkum} P.,  2019, \mn@doi [\apj]
  {10.3847/1538-4357/ab14f3}, \href
  {https://ui.adsabs.harvard.edu/abs/2019ApJ...875..155D} {875, 155}

\bibitem[\protect\citeauthoryear{{Danieli}, {van Dokkum}, {Conroy}, {Abraham}
  \& {Romanowsky}}{{Danieli} et~al.}{2019}]{Danieli.etal.2019}
{Danieli} S.,  {van Dokkum} P.,  {Conroy} C.,  {Abraham} R.,   {Romanowsky}
  A.~J.,  2019, \mn@doi [\apjl] {10.3847/2041-8213/ab0e8c}, \href
  {https://ui.adsabs.harvard.edu/abs/2019ApJ...874L..12D} {874, L12}

\bibitem[\protect\citeauthoryear{{Danieli} et~al.,}{{Danieli}
  et~al.}{2022}]{Danieli.etal.2022}
{Danieli} S.,  et~al., 2022, \mn@doi [\apjl] {10.3847/2041-8213/ac590a}, \href
  {https://ui.adsabs.harvard.edu/abs/2022ApJ...927L..28D} {927, L28}

\bibitem[\protect\citeauthoryear{{Di Cintio}, {Brook}, {Dutton}, {Macci{\`o}},
  {Stinson}  \& {Knebe}}{{Di Cintio} et~al.}{2014}]{DiCintio.etal.2014}
{Di Cintio} A.,  {Brook} C.~B.,  {Dutton} A.~A.,  {Macci{\`o}} A.~V.,
  {Stinson} G.~S.,   {Knebe} A.,  2014, \mn@doi [\mnras]
  {10.1093/mnras/stu729}, \href
  {https://ui.adsabs.harvard.edu/abs/2014MNRAS.441.2986D} {441, 2986}

\bibitem[\protect\citeauthoryear{{Di Cintio}, {Brook}, {Dutton}, {Macci{\`o}},
  {Obreja}  \& {Dekel}}{{Di Cintio} et~al.}{2017}]{DiCintio.etal.2017}
{Di Cintio} A.,  {Brook} C.~B.,  {Dutton} A.~A.,  {Macci{\`o}} A.~V.,  {Obreja}
  A.,   {Dekel} A.,  2017, \mn@doi [\mnras] {10.1093/mnrasl/slw210}, \href
  {https://ui.adsabs.harvard.edu/abs/2017MNRAS.466L...1D} {466, L1}

\bibitem[\protect\citeauthoryear{{Di Teodoro} et~al.,}{{Di Teodoro}
  et~al.}{2019}]{DiTeodoro.etal.2019}
{Di Teodoro} E.~M.,  et~al., 2019, \mn@doi [\mnras] {10.1093/mnras/sty3095},
  \href {https://ui.adsabs.harvard.edu/abs/2019MNRAS.483..392D} {483, 392}

\bibitem[\protect\citeauthoryear{{Disney}}{{Disney}}{1976}]{Disney.1976}
{Disney} M.~J.,  1976, \mn@doi [\nat] {10.1038/263573a0}, \href
  {https://ui.adsabs.harvard.edu/abs/1976Natur.263..573D} {263, 573}

\bibitem[\protect\citeauthoryear{{Doppel}, {Sales}, {Benavides}, {Toloba},
  {Peng}, {Nelson}  \& {Navarro}}{{Doppel} et~al.}{2024}]{Doppel.etal.2024}
{Doppel} J.~E.,  {Sales} L.~V.,  {Benavides} J.~A.,  {Toloba} E.,  {Peng}
  E.~W.,  {Nelson} D.,   {Navarro} J.~F.,  2024, \mn@doi [\mnras]
  {10.1093/mnras/stae647}, \href
  {https://ui.adsabs.harvard.edu/abs/2024MNRAS.529.1827D} {529, 1827}

\bibitem[\protect\citeauthoryear{{Downing} \& {Oman}}{{Downing} \&
  {Oman}}{2023}]{Downing.Oman.2023}
{Downing} E.~R.,  {Oman} K.~A.,  2023, \mn@doi [\mnras]
  {10.1093/mnras/stad868}, \href
  {https://ui.adsabs.harvard.edu/abs/2023MNRAS.522.3318D} {522, 3318}

\bibitem[\protect\citeauthoryear{{Driver}, {Liske}, {Cross}, {De Propris}  \&
  {Allen}}{{Driver} et~al.}{2005}]{Driver.etal.2005}
{Driver} S.~P.,  {Liske} J.,  {Cross} N.~J.~G.,  {De Propris} R.,   {Allen}
  P.~D.,  2005, \mn@doi [\mnras] {10.1111/j.1365-2966.2005.08990.x}, \href
  {https://ui.adsabs.harvard.edu/abs/2005MNRAS.360...81D} {360, 81}

\bibitem[\protect\citeauthoryear{{Emsellem} et~al.,}{{Emsellem}
  et~al.}{2019}]{Emsellem.etal.2019}
{Emsellem} E.,  et~al., 2019, \mn@doi [\aap] {10.1051/0004-6361/201834909},
  \href {https://ui.adsabs.harvard.edu/abs/2019A&A...625A..76E} {625, A76}

\bibitem[\protect\citeauthoryear{{Feldmann}}{{Feldmann}}{2013}]{Feldmann.2013}
{Feldmann} R.,  2013, \mn@doi [\mnras] {10.1093/mnras/stt851}, \href
  {https://ui.adsabs.harvard.edu/abs/2013MNRAS.433.1910F} {433, 1910}

\bibitem[\protect\citeauthoryear{{Fensch} et~al.,}{{Fensch}
  et~al.}{2019}]{Fensch2019}
{Fensch} J.,  et~al., 2019, \mn@doi [A\&A] {10.1051/0004-6361/201834911}, \href
  {https://ui.adsabs.harvard.edu/abs/2019A&A...625A..77F} {625, A77}

\bibitem[\protect\citeauthoryear{{Ferr{\'e}-Mateu} et~al.,}{{Ferr{\'e}-Mateu}
  et~al.}{2018}]{FerreMateu2018}
{Ferr{\'e}-Mateu} A.,  et~al., 2018, \mn@doi [MNRAS] {10.1093/mnras/sty1597},
  \href {https://ui.adsabs.harvard.edu/abs/2018MNRAS.479.4891F} {479, 4891}

\bibitem[\protect\citeauthoryear{{Ferr{\'e}-Mateu}, {Gannon}, {Forbes},
  {Buzzo}, {Romanowsky}  \& {Brodie}}{{Ferr{\'e}-Mateu}
  et~al.}{2023}]{FerreMateu.etal.2023}
{Ferr{\'e}-Mateu} A.,  {Gannon} J.~S.,  {Forbes} D.~A.,  {Buzzo} M.~L.,
  {Romanowsky} A.~J.,   {Brodie} J.~P.,  2023, \mn@doi [\mnras]
  {10.1093/mnras/stad3102}, \href
  {https://ui.adsabs.harvard.edu/abs/2023MNRAS.526.4735F} {526, 4735}

\bibitem[\protect\citeauthoryear{{Fielder}, {Jones}, {Sand}, {Bennet},
  {Crnojevic}, {Karunakaran}, {Mutlu-Pakdil}  \& {Spekkens}}{{Fielder}
  et~al.}{2024}]{Fielder.etal.2024}
{Fielder} C.,  {Jones} M.,  {Sand} D.,  {Bennet} P.,  {Crnojevic} D.,
  {Karunakaran} A.,  {Mutlu-Pakdil} B.,   {Spekkens} K.,  2024, \mn@doi [arXiv
  e-prints] {10.48550/arXiv.2401.01931}, \href
  {https://ui.adsabs.harvard.edu/abs/2024arXiv240101931F} {p. arXiv:2401.01931}

\bibitem[\protect\citeauthoryear{{Forbes} \& {Gannon}}{{Forbes} \&
  {Gannon}}{2024}]{Forbes.Gannon.2024}
{Forbes} D.~A.,  {Gannon} J.,  2024, \mn@doi [\mnras] {10.1093/mnras/stad4004},
  \href {https://ui.adsabs.harvard.edu/abs/2024MNRAS.528..608F} {528, 608}

\bibitem[\protect\citeauthoryear{{Forbes}, {Read}, {Gieles}  \&
  {Collins}}{{Forbes} et~al.}{2018a}]{Forbes.etal.2018}
{Forbes} D.~A.,  {Read} J.~I.,  {Gieles} M.,   {Collins} M. L.~M.,  2018a,
  \mn@doi [\mnras] {10.1093/mnras/sty2584}, \href
  {https://ui.adsabs.harvard.edu/abs/2018MNRAS.481.5592F} {481, 5592}

\bibitem[\protect\citeauthoryear{{Forbes}, {Read}, {Gieles}  \&
  {Collins}}{{Forbes} et~al.}{2018b}]{Forbes2018}
{Forbes} D.~A.,  {Read} J.~I.,  {Gieles} M.,   {Collins} M. L.~M.,  2018b,
  \mn@doi [MNRAS] {10.1093/mnras/sty2584}, \href
  {https://ui.adsabs.harvard.edu/abs/2018MNRAS.481.5592F} {481, 5592}

\bibitem[\protect\citeauthoryear{{Forbes}, {Alabi}, {Romanowsky}, {Brodie}  \&
  {Arimoto}}{{Forbes} et~al.}{2020}]{Forbes.etal.2020}
{Forbes} D.~A.,  {Alabi} A.,  {Romanowsky} A.~J.,  {Brodie} J.~P.,   {Arimoto}
  N.,  2020, \mn@doi [\mnras] {10.1093/mnras/staa180}, \href
  {https://ui.adsabs.harvard.edu/abs/2020MNRAS.492.4874F} {492, 4874}

\bibitem[\protect\citeauthoryear{{Forbes}, {Gannon}, {Romanowsky}, {Alabi},
  {Brodie}, {Couch}  \& {Ferr{\'e}-Mateu}}{{Forbes} et~al.}{2021}]{Forbes2021}
{Forbes} D.~A.,  {Gannon} J.~S.,  {Romanowsky} A.~J.,  {Alabi} A.,  {Brodie}
  J.~P.,  {Couch} W.~J.,   {Ferr{\'e}-Mateu} A.,  2021, \mn@doi [MNRAS]
  {10.1093/mnras/staa3289}, \href
  {https://ui.adsabs.harvard.edu/abs/2021MNRAS.500.1279F} {500, 1279}

\bibitem[\protect\citeauthoryear{{Forbes} et~al.,}{{Forbes}
  et~al.}{2023}]{Forbes.etal.2023}
{Forbes} D.~A.,  et~al., 2023, \mn@doi [\mnras] {10.1093/mnrasl/slad101}, \href
  {https://ui.adsabs.harvard.edu/abs/2023MNRAS.525L..93F} {525, L93}

\bibitem[\protect\citeauthoryear{{Gannon}, {Forbes}, {Romanowsky},
  {Ferr{\'e}-Mateu}, {Couch}  \& {Brodie}}{{Gannon} et~al.}{2020}]{Gannon2020}
{Gannon} J.~S.,  {Forbes} D.~A.,  {Romanowsky} A.~J.,  {Ferr{\'e}-Mateu} A.,
  {Couch} W.~J.,   {Brodie} J.~P.,  2020, \mn@doi [MNRAS]
  {10.1093/mnras/staa1282}, \href
  {https://ui.adsabs.harvard.edu/abs/2020MNRAS.495.2582G} {495, 2582}

\bibitem[\protect\citeauthoryear{{Gannon} et~al.,}{{Gannon}
  et~al.}{2021}]{Gannon2021}
{Gannon} J.~S.,  et~al., 2021, \mn@doi [MNRAS] {10.1093/mnras/stab277}, \href
  {https://ui.adsabs.harvard.edu/abs/2021MNRAS.502.3144G} {502, 3144}

\bibitem[\protect\citeauthoryear{{Gannon} et~al.,}{{Gannon}
  et~al.}{2022a}]{Gannon.etal.2022}
{Gannon} J.~S.,  et~al., 2022a, \mn@doi [\mnras] {10.1093/mnras/stab3297},
  \href {https://ui.adsabs.harvard.edu/abs/2022MNRAS.510..946G} {510, 946}

\bibitem[\protect\citeauthoryear{{Gannon} et~al.,}{{Gannon}
  et~al.}{2022b}]{Gannon2022}
{Gannon} J.~S.,  et~al., 2022b, \mn@doi [MNRAS] {10.1093/mnras/stab3297}, \href
  {https://ui.adsabs.harvard.edu/abs/2022MNRAS.510..946G} {510, 946}

\bibitem[\protect\citeauthoryear{{Gannon}, {Forbes}, {Brodie}, {Romanowsky},
  {Couch}  \& {Ferr{\'e}-Mateu}}{{Gannon} et~al.}{2023}]{Gannon.etal.2023}
{Gannon} J.~S.,  {Forbes} D.~A.,  {Brodie} J.~P.,  {Romanowsky} A.~J.,  {Couch}
  W.~J.,   {Ferr{\'e}-Mateu} A.,  2023, \mn@doi [MNRAS]
  {10.1093/mnras/stac3264}, \href
  {https://ui.adsabs.harvard.edu/abs/2023MNRAS.518.3653G} {518, 3653}

\bibitem[\protect\citeauthoryear{{Gannon}, {Ferr{\'e}-Mateu}, {Forbes},
  {Brodie}, {Buzzo}  \& {Romanowsky}}{{Gannon}
  et~al.}{2024}]{Gannon.etal.2024cat}
{Gannon} J.~S.,  {Ferr{\'e}-Mateu} A.,  {Forbes} D.~A.,  {Brodie} J.~P.,
  {Buzzo} M.~L.,   {Romanowsky} A.~J.,  2024, \mn@doi [arXiv e-prints]
  {10.48550/arXiv.2405.09104}, \href
  {https://ui.adsabs.harvard.edu/abs/2024arXiv240509104G} {p. arXiv:2405.09104}

\bibitem[\protect\citeauthoryear{{Garrison-Kimmel}, {Boylan-Kolchin}, {Bullock}
   \& {Lee}}{{Garrison-Kimmel} et~al.}{2014}]{Garrison_Kimmel.etal.2014}
{Garrison-Kimmel} S.,  {Boylan-Kolchin} M.,  {Bullock} J.~S.,   {Lee} K.,
  2014, \mn@doi [\mnras] {10.1093/mnras/stt2377}, \href
  {https://ui.adsabs.harvard.edu/abs/2014MNRAS.438.2578G} {438, 2578}

\bibitem[\protect\citeauthoryear{{Geller}, {Diaferio}, {Kurtz}, {Dell'Antonio}
  \& {Fabricant}}{{Geller} et~al.}{2012}]{Geller.etal.2012}
{Geller} M.~J.,  {Diaferio} A.,  {Kurtz} M.~J.,  {Dell'Antonio} I.~P.,
  {Fabricant} D.~G.,  2012, \mn@doi [\aj] {10.1088/0004-6256/143/4/102}, \href
  {https://ui.adsabs.harvard.edu/abs/2012AJ....143..102G} {143, 102}

\bibitem[\protect\citeauthoryear{{Georgiev}, {Puzia}, {Goudfrooij}  \&
  {Hilker}}{{Georgiev} et~al.}{2010}]{Georgiev.etal.2010}
{Georgiev} I.~Y.,  {Puzia} T.~H.,  {Goudfrooij} P.,   {Hilker} M.,  2010,
  \mn@doi [\mnras] {10.1111/j.1365-2966.2010.16802.x}, \href
  {https://ui.adsabs.harvard.edu/abs/2010MNRAS.406.1967G} {406, 1967}

\bibitem[\protect\citeauthoryear{{Golini}, {Montes}, {Carrasco}, {Rom{\'a}n}
  \& {Trujillo}}{{Golini} et~al.}{2024}]{Golini.etal.2024}
{Golini} G.,  {Montes} M.,  {Carrasco} E.~R.,  {Rom{\'a}n} J.,   {Trujillo} I.,
   2024, \mn@doi [\aap] {10.1051/0004-6361/202348300}, \href
  {https://ui.adsabs.harvard.edu/abs/2024A&A...684A..99G} {684, A99}

\bibitem[\protect\citeauthoryear{{Governato} et~al.,}{{Governato}
  et~al.}{2012}]{Governato.etal.2012}
{Governato} F.,  et~al., 2012, \mn@doi [\mnras]
  {10.1111/j.1365-2966.2012.20696.x}, \href
  {https://ui.adsabs.harvard.edu/abs/2012MNRAS.422.1231G} {422, 1231}

\bibitem[\protect\citeauthoryear{{Griffen}, {Ji}, {Dooley}, {G{\'o}mez},
  {Vogelsberger}, {O'Shea}  \& {Frebel}}{{Griffen}
  et~al.}{2016}]{Griffen.etal.2016}
{Griffen} B.~F.,  {Ji} A.~P.,  {Dooley} G.~A.,  {G{\'o}mez} F.~A.,
  {Vogelsberger} M.,  {O'Shea} B.~W.,   {Frebel} A.,  2016, \mn@doi [\apj]
  {10.3847/0004-637X/818/1/10}, \href
  {https://ui.adsabs.harvard.edu/abs/2016ApJ...818...10G} {818, 10}

\bibitem[\protect\citeauthoryear{{Gu} et~al.,}{{Gu} et~al.}{2018}]{Gu2018}
{Gu} M.,  et~al., 2018, \mn@doi [ApJ] {10.3847/1538-4357/aabbae}, \href
  {https://ui.adsabs.harvard.edu/abs/2018ApJ...859...37G} {859, 37}

\bibitem[\protect\citeauthoryear{{Guo} et~al.,}{{Guo}
  et~al.}{2020}]{Guo.etal.2020}
{Guo} Q.,  et~al., 2020, \mn@doi [Nature Astronomy]
  {10.1038/s41550-019-0930-9}, \href
  {https://ui.adsabs.harvard.edu/abs/2020NatAs...4..246G} {4, 246}

\bibitem[\protect\citeauthoryear{{Harris} \& {Zaritsky}}{{Harris} \&
  {Zaritsky}}{2004}]{Harris.Zaritsky.2004}
{Harris} J.,  {Zaritsky} D.,  2004, \mn@doi [\aj] {10.1086/381953}, \href
  {https://ui.adsabs.harvard.edu/abs/2004AJ....127.1531H} {127, 1531}

\bibitem[\protect\citeauthoryear{Harris et~al.,}{Harris et~al.}{2020}]{numpy}
Harris C.~R.,  et~al., 2020, \mn@doi [Nature] {10.1038/s41586-020-2649-2}, 585,
  357–362

\bibitem[\protect\citeauthoryear{Hunter}{Hunter}{2007}]{matplotlib}
Hunter J.~D.,  2007, \mn@doi [Computing In Science \& Engineering]
  {10.1109/MCSE.2007.55}, 9, 90

\bibitem[\protect\citeauthoryear{{Impey} \& {Bothun}}{{Impey} \&
  {Bothun}}{1997}]{Impey.Bothun.1997}
{Impey} C.,  {Bothun} G.,  1997, \mn@doi [\araa]
  {10.1146/annurev.astro.35.1.267}, \href
  {https://ui.adsabs.harvard.edu/abs/1997ARA&A..35..267I} {35, 267}

\bibitem[\protect\citeauthoryear{{Impey}, {Bothun}  \& {Malin}}{{Impey}
  et~al.}{1988}]{Impey.etal.1988}
{Impey} C.,  {Bothun} G.,   {Malin} D.,  1988, \mn@doi [\apj] {10.1086/166500},
  \href {https://ui.adsabs.harvard.edu/abs/1988ApJ...330..634I} {330, 634}

\bibitem[\protect\citeauthoryear{{Iodice} et~al.,}{{Iodice}
  et~al.}{2020}]{Iodice.etal.2020}
{Iodice} E.,  et~al., 2020, \mn@doi [A\&A] {10.1051/0004-6361/202038523}, \href
  {https://ui.adsabs.harvard.edu/abs/2020A&A...642A..48I} {642, A48}

\bibitem[\protect\citeauthoryear{{Iodice} et~al.,}{{Iodice}
  et~al.}{2023}]{Iodice.etal.2023}
{Iodice} E.,  et~al., 2023, \mn@doi [arXiv e-prints]
  {10.48550/arXiv.2308.11493}, \href
  {https://ui.adsabs.harvard.edu/abs/2023arXiv230811493I} {p. arXiv:2308.11493}

\bibitem[\protect\citeauthoryear{{Ivleva}, {Remus}, {Valenzuela}  \&
  {Dolag}}{{Ivleva} et~al.}{2024}]{Ivleva.etal.2024}
{Ivleva} A.,  {Remus} R.-S.,  {Valenzuela} L.~M.,   {Dolag} K.,  2024, \mn@doi
  [arXiv e-prints] {10.48550/arXiv.2402.09060}, \href
  {https://ui.adsabs.harvard.edu/abs/2024arXiv240209060I} {p. arXiv:2402.09060}

\bibitem[\protect\citeauthoryear{{Jackson} et~al.,}{{Jackson}
  et~al.}{2021}]{Jackson.etal.2021}
{Jackson} R.~A.,  et~al., 2021, \mn@doi [\mnras] {10.1093/mnras/stab077}, \href
  {https://ui.adsabs.harvard.edu/abs/2021MNRAS.502.4262J} {502, 4262}

\bibitem[\protect\citeauthoryear{{Janssens} et~al.,}{{Janssens}
  et~al.}{2022}]{Janssens.etal.2022}
{Janssens} S.~R.,  et~al., 2022, \mn@doi [MNRAS] {10.1093/mnras/stac2717},
  \href {https://ui.adsabs.harvard.edu/abs/2022MNRAS.517..858J} {517, 858}

\bibitem[\protect\citeauthoryear{{Jiang}, {Dekel}, {Freundlich}, {Romanowsky},
  {Dutton}, {Macci{\`o}}  \& {Di Cintio}}{{Jiang}
  et~al.}{2019}]{Jiang.etal.2019}
{Jiang} F.,  {Dekel} A.,  {Freundlich} J.,  {Romanowsky} A.~J.,  {Dutton}
  A.~A.,  {Macci{\`o}} A.~V.,   {Di Cintio} A.,  2019, \mn@doi [\mnras]
  {10.1093/mnras/stz1499}, \href
  {https://ui.adsabs.harvard.edu/abs/2019MNRAS.487.5272J} {487, 5272}

\bibitem[\protect\citeauthoryear{{Karachentsev}, {Makarova}, {Sharina}  \&
  {Karachentseva}}{{Karachentsev} et~al.}{2017}]{Karachentsev2017}
{Karachentsev} I.~D.,  {Makarova} L.~N.,  {Sharina} M.~E.,   {Karachentseva}
  V.~E.,  2017, \mn@doi [Astrophysical Bulletin] {10.1134/S1990341317040022},
  \href {https://ui.adsabs.harvard.edu/abs/2017AstBu..72..376K} {72, 376}

\bibitem[\protect\citeauthoryear{{Keim} et~al.,}{{Keim}
  et~al.}{2022}]{Keim.etal.2022}
{Keim} M.~A.,  et~al., 2022, \mn@doi [\apj] {10.3847/1538-4357/ac7dab}, \href
  {https://ui.adsabs.harvard.edu/abs/2022ApJ...935..160K} {935, 160}

\bibitem[\protect\citeauthoryear{{Kelley}, {Bullock}, {Garrison-Kimmel},
  {Boylan-Kolchin}, {Pawlowski}  \& {Graus}}{{Kelley}
  et~al.}{2019}]{Kelley.etal.2019}
{Kelley} T.,  {Bullock} J.~S.,  {Garrison-Kimmel} S.,  {Boylan-Kolchin} M.,
  {Pawlowski} M.~S.,   {Graus} A.~S.,  2019, \mn@doi [\mnras]
  {10.1093/mnras/stz1553}, \href
  {https://ui.adsabs.harvard.edu/abs/2019MNRAS.487.4409K} {487, 4409}

\bibitem[\protect\citeauthoryear{{Koda}, {Yagi}, {Yamanoi}  \&
  {Komiyama}}{{Koda} et~al.}{2015}]{Koda.etal.2015}
{Koda} J.,  {Yagi} M.,  {Yamanoi} H.,   {Komiyama} Y.,  2015, \mn@doi [\apjl]
  {10.1088/2041-8205/807/1/L2}, \href
  {https://ui.adsabs.harvard.edu/abs/2015ApJ...807L...2K} {807, L2}

\bibitem[\protect\citeauthoryear{{Kov{\'a}cs}, {Bogd{\'a}n}  \&
  {Canning}}{{Kov{\'a}cs} et~al.}{2019}]{Kovacs.etal.2019}
{Kov{\'a}cs} O.~E.,  {Bogd{\'a}n} {\'A}.,   {Canning} R. E.~A.,  2019, \mn@doi
  [\apjl] {10.3847/2041-8213/ab2916}, \href
  {https://ui.adsabs.harvard.edu/abs/2019ApJ...879L..12K} {879, L12}

\bibitem[\protect\citeauthoryear{{Kravtsov} \& {Belokurov}}{{Kravtsov} \&
  {Belokurov}}{2024}]{Kravtsov.Belokurov.2024}
{Kravtsov} A.,  {Belokurov} V.,  2024, \mn@doi [arXiv e-prints]
  {10.48550/arXiv.2405.04578}, \href
  {https://ui.adsabs.harvard.edu/abs/2024arXiv240504578K} {p. arXiv:2405.04578}

\bibitem[\protect\citeauthoryear{{Kravtsov} \& {Manwadkar}}{{Kravtsov} \&
  {Manwadkar}}{2022}]{Kravtsov.Manwadkar.2022}
{Kravtsov} A.,  {Manwadkar} V.,  2022, \mn@doi [\mnras]
  {10.1093/mnras/stac1439}, \href
  {https://ui.adsabs.harvard.edu/abs/2022MNRAS.514.2667K} {514, 2667}

\bibitem[\protect\citeauthoryear{{Kravtsov} \& {Wu}}{{Kravtsov} \&
  {Wu}}{2023}]{Kravtsov.Wu.2023}
{Kravtsov} A.,  {Wu} Z.,  2023, \mn@doi [\mnras] {10.1093/mnras/stad2219},
  \href {https://ui.adsabs.harvard.edu/abs/2023MNRAS.525..325K} {525, 325}

\bibitem[\protect\citeauthoryear{{Krumholz} \& {Dekel}}{{Krumholz} \&
  {Dekel}}{2012}]{Krumholz.Dekel.2012}
{Krumholz} M.~R.,  {Dekel} A.,  2012, \mn@doi [\apj]
  {10.1088/0004-637X/753/1/16}, \href
  {https://ui.adsabs.harvard.edu/abs/2012ApJ...753...16K} {753, 16}

\bibitem[\protect\citeauthoryear{{Laporte}, {Agnello}  \& {Navarro}}{{Laporte}
  et~al.}{2019}]{Laporte.etal.2019}
{Laporte} C. F.~P.,  {Agnello} A.,   {Navarro} J.~F.,  2019, \mn@doi [\mnras]
  {10.1093/mnras/sty2891}, \href
  {https://ui.adsabs.harvard.edu/abs/2019MNRAS.484..245L} {484, 245}

\bibitem[\protect\citeauthoryear{{Lazar} et~al.,}{{Lazar}
  et~al.}{2020}]{Lazar.etal.2020}
{Lazar} A.,  et~al., 2020, \mn@doi [\mnras] {10.1093/mnras/staa2101}, \href
  {https://ui.adsabs.harvard.edu/abs/2020MNRAS.497.2393L} {497, 2393}

\bibitem[\protect\citeauthoryear{{Lee}, {Shin}  \& {Kim}}{{Lee}
  et~al.}{2021}]{Lee.etal.2021}
{Lee} J.,  {Shin} E.-j.,   {Kim} J.-h.,  2021, \mn@doi [\apjl]
  {10.3847/2041-8213/ac16e0}, \href
  {https://ui.adsabs.harvard.edu/abs/2021ApJ...917L..15L} {917, L15}

\bibitem[\protect\citeauthoryear{{Lee}, {Shin}, {Kim}, {Shapiro}  \&
  {Chung}}{{Lee} et~al.}{2024}]{Lee.etal.2024}
{Lee} J.,  {Shin} E.-j.,  {Kim} J.-h.,  {Shapiro} P.~R.,   {Chung} E.,  2024,
  \mn@doi [\apj] {10.3847/1538-4357/ad2932}, \href
  {https://ui.adsabs.harvard.edu/abs/2024ApJ...966...72L} {966, 72}

\bibitem[\protect\citeauthoryear{{Leisman} et~al.,}{{Leisman}
  et~al.}{2017}]{Leisman.etal.2017}
{Leisman} L.,  et~al., 2017, \mn@doi [\apj] {10.3847/1538-4357/aa7575}, \href
  {https://ui.adsabs.harvard.edu/abs/2017ApJ...842..133L} {842, 133}

\bibitem[\protect\citeauthoryear{{Lewis}, {Brewer}  \& {Wan}}{{Lewis}
  et~al.}{2020}]{Lewis.etal.2020}
{Lewis} G.~F.,  {Brewer} B.~J.,   {Wan} Z.,  2020, \mn@doi [\mnras]
  {10.1093/mnrasl/slz157}, \href
  {https://ui.adsabs.harvard.edu/abs/2020MNRAS.491L...1L} {491, L1}

\bibitem[\protect\citeauthoryear{{Liao} et~al.,}{{Liao}
  et~al.}{2019}]{Liao.etal.2019}
{Liao} S.,  et~al., 2019, \mn@doi [\mnras] {10.1093/mnras/stz2969}, \href
  {https://ui.adsabs.harvard.edu/abs/2019MNRAS.490.5182L} {490, 5182}

\bibitem[\protect\citeauthoryear{{Lilly}, {Carollo}, {Pipino}, {Renzini}  \&
  {Peng}}{{Lilly} et~al.}{2013}]{Lilly.etal.2013}
{Lilly} S.~J.,  {Carollo} C.~M.,  {Pipino} A.,  {Renzini} A.,   {Peng} Y.,
  2013, \mn@doi [\apj] {10.1088/0004-637X/772/2/119}, \href
  {https://ui.adsabs.harvard.edu/abs/2013ApJ...772..119L} {772, 119}

\bibitem[\protect\citeauthoryear{{Lim}, {Peng}, {C{\^o}t{\'e}}, {Sales}, {den
  Brok}, {Blakeslee}  \& {Guhathakurta}}{{Lim} et~al.}{2018}]{Lim2018}
{Lim} S.,  {Peng} E.~W.,  {C{\^o}t{\'e}} P.,  {Sales} L.~V.,  {den Brok} M.,
  {Blakeslee} J.~P.,   {Guhathakurta} P.,  2018, \mn@doi [ApJ]
  {10.3847/1538-4357/aacb81}, \href
  {https://ui.adsabs.harvard.edu/abs/2018ApJ...862...82L} {862, 82}

\bibitem[\protect\citeauthoryear{{Lim} et~al.,}{{Lim} et~al.}{2020}]{Lim2020}
{Lim} S.,  et~al., 2020, \mn@doi [ApJ] {10.3847/1538-4357/aba433}, \href
  {https://ui.adsabs.harvard.edu/abs/2020ApJ...899...69L} {899, 69}

\bibitem[\protect\citeauthoryear{{Mancera Pi{\~n}a}, {Peletier}, {Aguerri},
  {Venhola}, {Trager}  \& {Choque Challapa}}{{Mancera Pi{\~n}a}
  et~al.}{2018}]{Mancera_Pina.etal.2018}
{Mancera Pi{\~n}a} P.~E.,  {Peletier} R.~F.,  {Aguerri} J.~A.~L.,  {Venhola}
  A.,  {Trager} S.,   {Choque Challapa} N.,  2018, \mn@doi [\mnras]
  {10.1093/mnras/sty2574}, \href
  {https://ui.adsabs.harvard.edu/abs/2018MNRAS.481.4381M} {481, 4381}

\bibitem[\protect\citeauthoryear{{Mancera Pi{\~n}a}, {Aguerri}, {Peletier},
  {Venhola}, {Trager}  \& {Choque Challapa}}{{Mancera Pi{\~n}a}
  et~al.}{2019}]{Mancera_Pina.etal.2019}
{Mancera Pi{\~n}a} P.~E.,  {Aguerri} J.~A.~L.,  {Peletier} R.~F.,  {Venhola}
  A.,  {Trager} S.,   {Choque Challapa} N.,  2019, \mn@doi [\mnras]
  {10.1093/mnras/stz238}, \href
  {https://ui.adsabs.harvard.edu/abs/2019MNRAS.485.1036M} {485, 1036}

\bibitem[\protect\citeauthoryear{{Mancera Pi{\~n}a} et~al.,}{{Mancera Pi{\~n}a}
  et~al.}{2020}]{Mancera_Pina.etal.2020}
{Mancera Pi{\~n}a} P.~E.,  et~al., 2020, \mn@doi [\mnras]
  {10.1093/mnras/staa1256}, \href
  {https://ui.adsabs.harvard.edu/abs/2020MNRAS.495.3636M} {495, 3636}

\bibitem[\protect\citeauthoryear{{Manwadkar} \& {Kravtsov}}{{Manwadkar} \&
  {Kravtsov}}{2022}]{Manwadkar.Kravtsov.2022}
{Manwadkar} V.,  {Kravtsov} A.~V.,  2022, \mn@doi [\mnras]
  {10.1093/mnras/stac2452}, \href
  {https://ui.adsabs.harvard.edu/abs/2022MNRAS.516.3944M} {516, 3944}

\bibitem[\protect\citeauthoryear{{Marleau} et~al.,}{{Marleau}
  et~al.}{2024}]{Marleau.etal.2024}
{Marleau} F.~R.,  et~al., 2024, \mn@doi [arXiv e-prints]
  {10.48550/arXiv.2405.13502}, \href
  {https://ui.adsabs.harvard.edu/abs/2024arXiv240513502M} {p. arXiv:2405.13502}

\bibitem[\protect\citeauthoryear{{Mart{\'\i}n-Navarro}
  et~al.,}{{Mart{\'\i}n-Navarro} et~al.}{2019}]{MartinNavarro2019}
{Mart{\'\i}n-Navarro} I.,  et~al., 2019, \mn@doi [MNRAS]
  {10.1093/mnras/stz252}, \href
  {https://ui.adsabs.harvard.edu/abs/2019MNRAS.484.3425M} {484, 3425}

\bibitem[\protect\citeauthoryear{{Martin} et~al.,}{{Martin}
  et~al.}{2016}]{Martin2016}
{Martin} N.~F.,  et~al., 2016, \mn@doi [ApJ] {10.3847/1538-4357/833/2/167},
  \href {https://ui.adsabs.harvard.edu/abs/2016ApJ...833..167M} {833, 167}

\bibitem[\protect\citeauthoryear{{Martin}, {Collins}, {Longeard}  \&
  {Tollerud}}{{Martin} et~al.}{2018}]{Martin.etal.2018}
{Martin} N.~F.,  {Collins} M. L.~M.,  {Longeard} N.,   {Tollerud} E.,  2018,
  \mn@doi [\apjl] {10.3847/2041-8213/aac216}, \href
  {https://ui.adsabs.harvard.edu/abs/2018ApJ...859L...5M} {859, L5}

\bibitem[\protect\citeauthoryear{{Martin} et~al.,}{{Martin}
  et~al.}{2019}]{Martin.etal.2019}
{Martin} G.,  et~al., 2019, \mn@doi [\mnras] {10.1093/mnras/stz356}, \href
  {https://ui.adsabs.harvard.edu/abs/2019MNRAS.485..796M} {485, 796}

\bibitem[\protect\citeauthoryear{{Mart{\'\i}nez-Delgado}
  et~al.,}{{Mart{\'\i}nez-Delgado} et~al.}{2016}]{MartinezDelgado2016}
{Mart{\'\i}nez-Delgado} D.,  et~al., 2016, \mn@doi [AJ]
  {10.3847/0004-6256/151/4/96}, \href
  {https://ui.adsabs.harvard.edu/abs/2016AJ....151...96M} {151, 96}

\bibitem[\protect\citeauthoryear{{McConnachie}}{{McConnachie}}{2012}]{mcconnachie2012}
{McConnachie} A.~W.,  2012, \mn@doi [AJ] {10.1088/0004-6256/144/1/4}, \href
  {https://ui.adsabs.harvard.edu/abs/2012AJ....144....4M} {144, 4}

\bibitem[\protect\citeauthoryear{{McGaugh}, {Bothun}  \& {Schombert}}{{McGaugh}
  et~al.}{1995}]{McGaugh.etal.1995}
{McGaugh} S.~S.,  {Bothun} G.~D.,   {Schombert} J.~M.,  1995, \mn@doi [\aj]
  {10.1086/117543}, \href
  {https://ui.adsabs.harvard.edu/abs/1995AJ....110..573M} {110, 573}

\bibitem[\protect\citeauthoryear{{Mihos} et~al.,}{{Mihos}
  et~al.}{2022}]{Mihos2022}
{Mihos} J.~C.,  et~al., 2022, \mn@doi [ApJ] {10.3847/1538-4357/ac35d9}, \href
  {https://ui.adsabs.harvard.edu/abs/2022ApJ...924...87M} {924, 87}

\bibitem[\protect\citeauthoryear{{Montes} et~al.,}{{Montes}
  et~al.}{2024}]{Montes.etal.2024}
{Montes} M.,  et~al., 2024, \mn@doi [\aap] {10.1051/0004-6361/202347667}, \href
  {https://ui.adsabs.harvard.edu/abs/2024A&A...681A..15M} {681, A15}

\bibitem[\protect\citeauthoryear{{Moreno} et~al.,}{{Moreno}
  et~al.}{2022}]{Moreno.etal.2022}
{Moreno} J.,  et~al., 2022, \mn@doi [Nature Astronomy]
  {10.1038/s41550-021-01598-4}, \href
  {https://ui.adsabs.harvard.edu/abs/2022NatAs...6..496M} {6, 496}

\bibitem[\protect\citeauthoryear{{Mu{\~n}oz}, {C{\^o}t{\'e}}, {Santana},
  {Geha}, {Simon}, {Oyarz{\'u}n}, {Stetson}  \& {Djorgovski}}{{Mu{\~n}oz}
  et~al.}{2018}]{Munoz.etal.2018}
{Mu{\~n}oz} R.~R.,  {C{\^o}t{\'e}} P.,  {Santana} F.~A.,  {Geha} M.,  {Simon}
  J.~D.,  {Oyarz{\'u}n} G.~A.,  {Stetson} P.~B.,   {Djorgovski} S.~G.,  2018,
  \mn@doi [\apj] {10.3847/1538-4357/aac16b}, \href
  {https://ui.adsabs.harvard.edu/abs/2018ApJ...860...66M} {860, 66}

\bibitem[\protect\citeauthoryear{{M{\"u}ller} et~al.,}{{M{\"u}ller}
  et~al.}{2020}]{Muller2020}
{M{\"u}ller} O.,  et~al., 2020, \mn@doi [A\&A] {10.1051/0004-6361/202038351},
  \href {https://ui.adsabs.harvard.edu/abs/2020A&A...640A.106M} {640, A106}

\bibitem[\protect\citeauthoryear{{M{\"u}ller} et~al.,}{{M{\"u}ller}
  et~al.}{2021}]{Muller2021}
{M{\"u}ller} O.,  et~al., 2021, \mn@doi [ApJ] {10.3847/1538-4357/ac2831}, \href
  {https://ui.adsabs.harvard.edu/abs/2021ApJ...923....9M} {923, 9}

\bibitem[\protect\citeauthoryear{{Nandi}, {Banerjee}  \& {Narayanan}}{{Nandi}
  et~al.}{2023}]{Nandi.etal.2023}
{Nandi} N.,  {Banerjee} A.,   {Narayanan} G.,  2023, \mn@doi [arXiv e-prints]
  {10.48550/arXiv.2310.08925}, \href
  {https://ui.adsabs.harvard.edu/abs/2023arXiv231008925N} {p. arXiv:2310.08925}

\bibitem[\protect\citeauthoryear{{Navarro}, {Frenk}  \& {White}}{{Navarro}
  et~al.}{1997}]{Navarro.etal.1997}
{Navarro} J.~F.,  {Frenk} C.~S.,   {White} S. D.~M.,  1997, \mn@doi [\apj]
  {10.1086/304888}, \href
  {https://ui.adsabs.harvard.edu/abs/1997ApJ...490..493N} {490, 493}

\bibitem[\protect\citeauthoryear{Oliphant}{Oliphant}{2015}]{NumPy2015}
Oliphant T.~E.,  2015, Guide to NumPy, 2nd edn.
CreateSpace Independent Publishing Platform, USA

\bibitem[\protect\citeauthoryear{{Oman}, {Marasco}, {Navarro}, {Frenk},
  {Schaye}  \& {Ben{\'\i}tez-Llambay}}{{Oman} et~al.}{2019}]{Oman.etal.2019}
{Oman} K.~A.,  {Marasco} A.,  {Navarro} J.~F.,  {Frenk} C.~S.,  {Schaye} J.,
  {Ben{\'\i}tez-Llambay} A.,  2019, \mn@doi [\mnras] {10.1093/mnras/sty2687},
  \href {https://ui.adsabs.harvard.edu/abs/2019MNRAS.482..821O} {482, 821}

\bibitem[\protect\citeauthoryear{{Pontzen} \& {Governato}}{{Pontzen} \&
  {Governato}}{2012}]{Pontzen.Governato.2012}
{Pontzen} A.,  {Governato} F.,  2012, \mn@doi [\mnras]
  {10.1111/j.1365-2966.2012.20571.x}, \href
  {https://ui.adsabs.harvard.edu/abs/2012MNRAS.421.3464P} {421, 3464}

\bibitem[\protect\citeauthoryear{{Prole} et~al.,}{{Prole}
  et~al.}{2019}]{Prole.etal.2019}
{Prole} D.~J.,  et~al., 2019, \mn@doi [\mnras] {10.1093/mnras/stz326}, \href
  {https://ui.adsabs.harvard.edu/abs/2019MNRAS.484.4865P} {484, 4865}

\bibitem[\protect\citeauthoryear{{Read}, {Agertz}  \& {Collins}}{{Read}
  et~al.}{2016a}]{Read.etal.2016}
{Read} J.~I.,  {Agertz} O.,   {Collins} M.~L.~M.,  2016a, \mn@doi [\mnras]
  {10.1093/mnras/stw713}, \href
  {https://ui.adsabs.harvard.edu/abs/2016MNRAS.459.2573R} {459, 2573}

\bibitem[\protect\citeauthoryear{{Read}, {Iorio}, {Agertz}  \&
  {Fraternali}}{{Read} et~al.}{2016b}]{Read.etal.2016b}
{Read} J.~I.,  {Iorio} G.,  {Agertz} O.,   {Fraternali} F.,  2016b, \mn@doi
  [\mnras] {10.1093/mnras/stw1876}, \href
  {https://ui.adsabs.harvard.edu/abs/2016MNRAS.462.3628R} {462, 3628}

\bibitem[\protect\citeauthoryear{{Rom{\'a}n} \& {Trujillo}}{{Rom{\'a}n} \&
  {Trujillo}}{2017}]{Roman.Trugillo.2017}
{Rom{\'a}n} J.,  {Trujillo} I.,  2017, \mn@doi [\mnras] {10.1093/mnras/stx438},
  \href {https://ui.adsabs.harvard.edu/abs/2017MNRAS.468..703R} {468, 703}

\bibitem[\protect\citeauthoryear{{Rong}, {Guo}, {Gao}, {Liao}, {Xie}, {Puzia},
  {Sun}  \& {Pan}}{{Rong} et~al.}{2017}]{Rong.etal.2017}
{Rong} Y.,  {Guo} Q.,  {Gao} L.,  {Liao} S.,  {Xie} L.,  {Puzia} T.~H.,  {Sun}
  S.,   {Pan} J.,  2017, \mn@doi [\mnras] {10.1093/mnras/stx1440}, \href
  {https://ui.adsabs.harvard.edu/abs/2017MNRAS.470.4231R} {470, 4231}

\bibitem[\protect\citeauthoryear{{Rong}, {Hu}, {He}, {Du}, {Guo}, {Wang},
  {Zhang}  \& {Mo}}{{Rong} et~al.}{2024}]{Rong.etal.2024}
{Rong} Y.,  {Hu} H.,  {He} M.,  {Du} W.,  {Guo} Q.,  {Wang} H.-Y.,  {Zhang}
  H.-X.,   {Mo} H.,  2024, \mn@doi [arXiv e-prints]
  {10.48550/arXiv.2404.00555}, \href
  {https://ui.adsabs.harvard.edu/abs/2024arXiv240400555R} {p. arXiv:2404.00555}

\bibitem[\protect\citeauthoryear{{Roper}, {Oman}, {Frenk},
  {Ben{\'\i}tez-Llambay}, {Navarro}  \& {Santos-Santos}}{{Roper}
  et~al.}{2023}]{Roper.etal.2023}
{Roper} F.~A.,  {Oman} K.~A.,  {Frenk} C.~S.,  {Ben{\'\i}tez-Llambay} A.,
  {Navarro} J.~F.,   {Santos-Santos} I. M.~E.,  2023, \mn@doi [\mnras]
  {10.1093/mnras/stad549}, \href
  {https://ui.adsabs.harvard.edu/abs/2023MNRAS.521.1316R} {521, 1316}

\bibitem[\protect\citeauthoryear{{Ruiz-Lara} et~al.,}{{Ruiz-Lara}
  et~al.}{2018}]{RuizLara2018}
{Ruiz-Lara} T.,  et~al., 2018, \mn@doi [MNRAS] {10.1093/mnras/sty1112}, \href
  {https://ui.adsabs.harvard.edu/abs/2018MNRAS.478.2034R} {478, 2034}

\bibitem[\protect\citeauthoryear{{Safarzadeh} \& {Scannapieco}}{{Safarzadeh} \&
  {Scannapieco}}{2017}]{Safarzadeh.Scannapieco.2017}
{Safarzadeh} M.,  {Scannapieco} E.,  2017, \mn@doi [\apj]
  {10.3847/1538-4357/aa94c8}, \href
  {https://ui.adsabs.harvard.edu/abs/2017ApJ...850...99S} {850, 99}

\bibitem[\protect\citeauthoryear{{Saifollahi}, {Zaritsky}, {Trujillo},
  {Peletier}, {Knapen}, {Amorisco}, {Beasley}  \& {Donnerstein}}{{Saifollahi}
  et~al.}{2022}]{Saifollahi2022}
{Saifollahi} T.,  {Zaritsky} D.,  {Trujillo} I.,  {Peletier} R.~F.,  {Knapen}
  J.~H.,  {Amorisco} N.,  {Beasley} M.~A.,   {Donnerstein} R.,  2022, \mn@doi
  [MNRAS] {10.1093/mnras/stac328}, \href
  {https://ui.adsabs.harvard.edu/abs/2022MNRAS.tmp..336S} {}

\bibitem[\protect\citeauthoryear{{Sales}, {Navarro}, {Pe{\~n}afiel}, {Peng},
  {Lim}  \& {Hernquist}}{{Sales} et~al.}{2020}]{Sales.etal.2020}
{Sales} L.~V.,  {Navarro} J.~F.,  {Pe{\~n}afiel} L.,  {Peng} E.~W.,  {Lim} S.,
   {Hernquist} L.,  2020, \mn@doi [\mnras] {10.1093/mnras/staa854}, \href
  {https://ui.adsabs.harvard.edu/abs/2020MNRAS.494.1848S} {494, 1848}

\bibitem[\protect\citeauthoryear{{Sandage} \& {Binggeli}}{{Sandage} \&
  {Binggeli}}{1984}]{Sandage.Bingelli.1984}
{Sandage} A.,  {Binggeli} B.,  1984, \mn@doi [\aj] {10.1086/113588}, \href
  {https://ui.adsabs.harvard.edu/abs/1984AJ.....89..919S} {89, 919}

\bibitem[\protect\citeauthoryear{{Sands} et~al.,}{{Sands}
  et~al.}{2024}]{Sands.etal.2024}
{Sands} I.~S.,  et~al., 2024, \mn@doi [arXiv e-prints]
  {10.48550/arXiv.2404.16247}, \href
  {https://ui.adsabs.harvard.edu/abs/2024arXiv240416247S} {p. arXiv:2404.16247}

\bibitem[\protect\citeauthoryear{{Schombert}, {Bothun}, {Schneider}  \&
  {McGaugh}}{{Schombert} et~al.}{1992}]{Schombert.etal.1992}
{Schombert} J.~M.,  {Bothun} G.~D.,  {Schneider} S.~E.,   {McGaugh} S.~S.,
  1992, \mn@doi [\aj] {10.1086/116129}, \href
  {https://ui.adsabs.harvard.edu/abs/1992AJ....103.1107S} {103, 1107}

\bibitem[\protect\citeauthoryear{{Shen}, {van Dokkum}  \& {Danieli}}{{Shen}
  et~al.}{2021a}]{Shen.etal.2021}
{Shen} Z.,  {van Dokkum} P.,   {Danieli} S.,  2021a, \mn@doi [\apj]
  {10.3847/1538-4357/abdd29}, \href
  {https://ui.adsabs.harvard.edu/abs/2021ApJ...909..179S} {909, 179}

\bibitem[\protect\citeauthoryear{{Shen} et~al.,}{{Shen}
  et~al.}{2021b}]{Shen.etal.2021b}
{Shen} Z.,  et~al., 2021b, \mn@doi [ApJL] {10.3847/2041-8213/ac0335}, \href
  {https://ui.adsabs.harvard.edu/abs/2021ApJ...914L..12S} {914, L12}

\bibitem[\protect\citeauthoryear{{Shen}, {van Dokkum}  \& {Danieli}}{{Shen}
  et~al.}{2023}]{Shen.etal.2023}
{Shen} Z.,  {van Dokkum} P.,   {Danieli} S.,  2023, \mn@doi [\apj]
  {10.3847/1538-4357/acfa70}, \href
  {https://ui.adsabs.harvard.edu/abs/2023ApJ...957....6S} {957, 6}

\bibitem[\protect\citeauthoryear{{Sif{\'o}n}, {van der Burg}, {Hoekstra},
  {Muzzin}  \& {Herbonnet}}{{Sif{\'o}n} et~al.}{2018}]{Sifon.etal.2018}
{Sif{\'o}n} C.,  {van der Burg} R. F.~J.,  {Hoekstra} H.,  {Muzzin} A.,
  {Herbonnet} R.,  2018, \mn@doi [\mnras] {10.1093/mnras/stx2648}, \href
  {https://ui.adsabs.harvard.edu/abs/2018MNRAS.473.3747S} {473, 3747}

\bibitem[\protect\citeauthoryear{{Skibba} et~al.,}{{Skibba}
  et~al.}{2012}]{Skibba.etal.2012}
{Skibba} R.~A.,  et~al., 2012, \mn@doi [\apj] {10.1088/0004-637X/761/1/42},
  \href {https://ui.adsabs.harvard.edu/abs/2012ApJ...761...42S} {761, 42}

\bibitem[\protect\citeauthoryear{{Somerville} et~al.,}{{Somerville}
  et~al.}{2018}]{Somerville.etal.2018}
{Somerville} R.~S.,  et~al., 2018, \mn@doi [\mnras] {10.1093/mnras/stx2040},
  \href {https://ui.adsabs.harvard.edu/abs/2018MNRAS.473.2714S} {473, 2714}

\bibitem[\protect\citeauthoryear{{Sprayberry}, {Impey}, {Bothun}  \&
  {Irwin}}{{Sprayberry} et~al.}{1995}]{Sprayberry.etal.1995}
{Sprayberry} D.,  {Impey} C.~D.,  {Bothun} G.~D.,   {Irwin} M.~J.,  1995,
  \mn@doi [\aj] {10.1086/117300}, \href
  {https://ui.adsabs.harvard.edu/abs/1995AJ....109..558S} {109, 558}

\bibitem[\protect\citeauthoryear{{Suess}, {Kriek}, {Price}  \& {Barro}}{{Suess}
  et~al.}{2019}]{Suess.etal.2019}
{Suess} K.~A.,  {Kriek} M.,  {Price} S.~H.,   {Barro} G.,  2019, \mn@doi
  [\apjl] {10.3847/2041-8213/ab4db3}, \href
  {https://ui.adsabs.harvard.edu/abs/2019ApJ...885L..22S} {885, L22}

\bibitem[\protect\citeauthoryear{{Tollet} et~al.,}{{Tollet}
  et~al.}{2016}]{Tollet.etal.2016}
{Tollet} E.,  et~al., 2016, \mn@doi [\mnras] {10.1093/mnras/stv2856}, \href
  {https://ui.adsabs.harvard.edu/abs/2016MNRAS.456.3542T} {456, 3542}

\bibitem[\protect\citeauthoryear{{Toloba} et~al.,}{{Toloba}
  et~al.}{2018}]{Toloba2018}
{Toloba} E.,  et~al., 2018, \mn@doi [ApJL] {10.3847/2041-8213/aab603}, \href
  {https://ui.adsabs.harvard.edu/abs/2018ApJ...856L..31T} {856, L31}

\bibitem[\protect\citeauthoryear{{Toloba} et~al.,}{{Toloba}
  et~al.}{2023}]{Toloba.etal.2023}
{Toloba} E.,  et~al., 2023, \mn@doi [arXiv e-prints]
  {10.48550/arXiv.2305.06369}, \href
  {https://ui.adsabs.harvard.edu/abs/2023arXiv230506369T} {p. arXiv:2305.06369}

\bibitem[\protect\citeauthoryear{{Torrealba} et~al.,}{{Torrealba}
  et~al.}{2019}]{Torrealba.etal.2019}
{Torrealba} G.,  et~al., 2019, \mn@doi [MNRAS] {10.1093/mnras/stz1624}, \href
  {https://ui.adsabs.harvard.edu/abs/2019MNRAS.488.2743T} {488, 2743}

\bibitem[\protect\citeauthoryear{{Tremmel}, {Wright}, {Brooks}, {Munshi},
  {Nagai}  \& {Quinn}}{{Tremmel} et~al.}{2020}]{Tremmel.etal.2020}
{Tremmel} M.,  {Wright} A.~C.,  {Brooks} A.~M.,  {Munshi} F.,  {Nagai} D.,
  {Quinn} T.~R.,  2020, \mn@doi [\mnras] {10.1093/mnras/staa2015}, \href
  {https://ui.adsabs.harvard.edu/abs/2020MNRAS.497.2786T} {497, 2786}

\bibitem[\protect\citeauthoryear{{Trujillo}}{{Trujillo}}{2021}]{Trujillo.2021}
{Trujillo} I.,  2021, \mn@doi [Nature Astronomy] {10.1038/s41550-021-01388-y},
  \href {https://ui.adsabs.harvard.edu/abs/2021NatAs...5.1182T} {5, 1182}

\bibitem[\protect\citeauthoryear{{Trujillo-Gomez}, {Kruijssen}, {Keller}  \&
  {Reina-Campos}}{{Trujillo-Gomez} et~al.}{2021}]{Trujillo_Gomez.etal.2021}
{Trujillo-Gomez} S.,  {Kruijssen} J.~M.~D.,  {Keller} B.~W.,   {Reina-Campos}
  M.,  2021, \mn@doi [\mnras] {10.1093/mnras/stab1895}, \href
  {https://ui.adsabs.harvard.edu/abs/2021MNRAS.506.4841T} {506, 4841}

\bibitem[\protect\citeauthoryear{{Trujillo-Gomez}, {Kruijssen}  \&
  {Reina-Campos}}{{Trujillo-Gomez} et~al.}{2022}]{Trujillo_Gomez.etal.2022}
{Trujillo-Gomez} S.,  {Kruijssen} J.~M.~D.,   {Reina-Campos} M.,  2022, \mn@doi
  [\mnras] {10.1093/mnras/stab3401}, \href
  {https://ui.adsabs.harvard.edu/abs/2022MNRAS.510.3356T} {510, 3356}

\bibitem[\protect\citeauthoryear{{Van Nest}, {Munshi}, {Wright}, {Tremmel},
  {Brooks}, {Nagai}  \& {Quinn}}{{Van Nest} et~al.}{2022}]{vanNest.etal.2022}
{Van Nest} J.~D.,  {Munshi} F.,  {Wright} A.~C.,  {Tremmel} M.,  {Brooks}
  A.~M.,  {Nagai} D.,   {Quinn} T.,  2022, \mn@doi [\apj]
  {10.3847/1538-4357/ac43b7}, \href
  {https://ui.adsabs.harvard.edu/abs/2022ApJ...926...92V} {926, 92}

\bibitem[\protect\citeauthoryear{{Venhola} et~al.,}{{Venhola}
  et~al.}{2017}]{Venhola.etal.2017}
{Venhola} A.,  et~al., 2017, \mn@doi [\aap] {10.1051/0004-6361/201730696},
  \href {https://ui.adsabs.harvard.edu/abs/2017A&A...608A.142V} {608, A142}

\bibitem[\protect\citeauthoryear{{Venhola} et~al.,}{{Venhola}
  et~al.}{2022}]{Venhola.etal.2022}
{Venhola} A.,  et~al., 2022, \mn@doi [\aap] {10.1051/0004-6361/202141756},
  \href {https://ui.adsabs.harvard.edu/abs/2022A&A...662A..43V} {662, A43}

\bibitem[\protect\citeauthoryear{{Villaume} et~al.,}{{Villaume}
  et~al.}{2022}]{Villaume2022}
{Villaume} A.,  et~al., 2022, \mn@doi [ApJ] {10.3847/1538-4357/ac341e}, \href
  {https://ui.adsabs.harvard.edu/abs/2022ApJ...924...32V} {924, 32}

\bibitem[\protect\citeauthoryear{Virtanen et~al.,}{Virtanen
  et~al.}{2020}]{SciPy}
Virtanen P.,  et~al., 2020, \mn@doi [Nature Methods]
  {10.1038/s41592-019-0686-2}, \href {https://rdcu.be/b08Wh} {17, 261}

\bibitem[\protect\citeauthoryear{{Wang}, {Cooper}, {Bose}, {Frenk}  \&
  {Hellwing}}{{Wang} et~al.}{2023}]{Wang.etal.2023ghostly}
{Wang} C.-W.,  {Cooper} A.~P.,  {Bose} S.,  {Frenk} C.~S.,   {Hellwing} W.~A.,
  2023, \mn@doi [\apj] {10.3847/1538-4357/ad011d}, \href
  {https://ui.adsabs.harvard.edu/abs/2023ApJ...958..166W} {958, 166}

\bibitem[\protect\citeauthoryear{{Webb} et~al.,}{{Webb}
  et~al.}{2022}]{Webb2022}
{Webb} K.~A.,  et~al., 2022, \mn@doi [MNRAS] {10.1093/mnras/stac2417}, \href
  {https://ui.adsabs.harvard.edu/abs/2022MNRAS.516.3318W} {516, 3318}

\bibitem[\protect\citeauthoryear{{Willmer}}{{Willmer}}{2018}]{Willmer.2018}
{Willmer} C. N.~A.,  2018, \mn@doi [\apjs] {10.3847/1538-4365/aabfdf}, \href
  {https://ui.adsabs.harvard.edu/abs/2018ApJS..236...47W} {236, 47}

\bibitem[\protect\citeauthoryear{{Wolf}, {Martinez}, {Bullock}, {Kaplinghat},
  {Geha}, {Mu{\~n}oz}, {Simon}  \& {Avedo}}{{Wolf}
  et~al.}{2010}]{Wolf.etal.2010}
{Wolf} J.,  {Martinez} G.~D.,  {Bullock} J.~S.,  {Kaplinghat} M.,  {Geha} M.,
  {Mu{\~n}oz} R.~R.,  {Simon} J.~D.,   {Avedo} F.~F.,  2010, \mn@doi [\mnras]
  {10.1111/j.1365-2966.2010.16753.x}, \href
  {https://ui.adsabs.harvard.edu/abs/2010MNRAS.406.1220W} {406, 1220}

\bibitem[\protect\citeauthoryear{{Wright}, {Tremmel}, {Brooks}, {Munshi},
  {Nagai}, {Sharma}  \& {Quinn}}{{Wright} et~al.}{2021}]{Wright.etal.2021}
{Wright} A.~C.,  {Tremmel} M.,  {Brooks} A.~M.,  {Munshi} F.,  {Nagai} D.,
  {Sharma} R.~S.,   {Quinn} T.~R.,  2021, \mn@doi [\mnras]
  {10.1093/mnras/stab081}, \href
  {https://ui.adsabs.harvard.edu/abs/2021MNRAS.502.5370W} {502, 5370}

\bibitem[\protect\citeauthoryear{{Wu} \& {Kravtsov}}{{Wu} \&
  {Kravtsov}}{2024}]{Wu.Kravtsov.2024}
{Wu} Z.,  {Kravtsov} A.,  2024, \mn@doi [arXiv e-prints]
  {10.48550/arXiv.2405.08066}, \href
  {https://ui.adsabs.harvard.edu/abs/2024arXiv240508066W} {p. arXiv:2405.08066}

\bibitem[\protect\citeauthoryear{{Yagi}, {Koda}, {Komiyama}  \&
  {Yamanoi}}{{Yagi} et~al.}{2016a}]{Yagi.etal.2016}
{Yagi} M.,  {Koda} J.,  {Komiyama} Y.,   {Yamanoi} H.,  2016a, \mn@doi [\apjs]
  {10.3847/0067-0049/225/1/11}, \href
  {https://ui.adsabs.harvard.edu/abs/2016ApJS..225...11Y} {225, 11}

\bibitem[\protect\citeauthoryear{{Yagi}, {Koda}, {Komiyama}  \&
  {Yamanoi}}{{Yagi} et~al.}{2016b}]{Yagi2016}
{Yagi} M.,  {Koda} J.,  {Komiyama} Y.,   {Yamanoi} H.,  2016b, \mn@doi [ApJS]
  {10.3847/0067-0049/225/1/11}, \href
  {https://ui.adsabs.harvard.edu/abs/2016ApJS..225...11Y} {225, 11}

\bibitem[\protect\citeauthoryear{{Yozin} \& {Bekki}}{{Yozin} \&
  {Bekki}}{2015}]{Yozin.Bekki.2015}
{Yozin} C.,  {Bekki} K.,  2015, \mn@doi [\mnras] {10.1093/mnras/stv1073}, \href
  {https://ui.adsabs.harvard.edu/abs/2015MNRAS.452..937Y} {452, 937}

\bibitem[\protect\citeauthoryear{{Z{\"o}ller}, {Kluge}, {Staiger}  \&
  {Bender}}{{Z{\"o}ller} et~al.}{2024}]{Zoeller.etal.2024}
{Z{\"o}ller} R.,  {Kluge} M.,  {Staiger} B.,   {Bender} R.,  2024, \mn@doi
  [\apjs] {10.3847/1538-4365/ad2775}, \href
  {https://ui.adsabs.harvard.edu/abs/2024ApJS..271...52Z} {271, 52}

\bibitem[\protect\citeauthoryear{{van Dokkum}, {Abraham}, {Merritt}, {Zhang},
  {Geha}  \& {Conroy}}{{van Dokkum} et~al.}{2015a}]{vanDokkum.etal.2015}
{van Dokkum} P.~G.,  {Abraham} R.,  {Merritt} A.,  {Zhang} J.,  {Geha} M.,
  {Conroy} C.,  2015a, \mn@doi [\apjl] {10.1088/2041-8205/798/2/L45}, \href
  {https://ui.adsabs.harvard.edu/abs/2015ApJ...798L..45V} {798, L45}

\bibitem[\protect\citeauthoryear{{van Dokkum} et~al.,}{{van Dokkum}
  et~al.}{2015b}]{vanDokkum.etal.2015b}
{van Dokkum} P.~G.,  et~al., 2015b, \mn@doi [\apjl]
  {10.1088/2041-8205/804/1/L26}, \href
  {https://ui.adsabs.harvard.edu/abs/2015ApJ...804L..26V} {804, L26}

\bibitem[\protect\citeauthoryear{{van Dokkum} et~al.,}{{van Dokkum}
  et~al.}{2016}]{vanDokkum2016}
{van Dokkum} P.,  et~al., 2016, \mn@doi [ApJL] {10.3847/2041-8205/828/1/L6},
  \href {https://ui.adsabs.harvard.edu/abs/2016ApJ...828L...6V} {828, L6}

\bibitem[\protect\citeauthoryear{{van Dokkum} et~al.,}{{van Dokkum}
  et~al.}{2017}]{vanDokkum.etal.2017}
{van Dokkum} P.,  et~al., 2017, \mn@doi [\apjl] {10.3847/2041-8213/aa7ca2},
  \href {https://ui.adsabs.harvard.edu/abs/2017ApJ...844L..11V} {844, L11}

\bibitem[\protect\citeauthoryear{{van Dokkum} et~al.,}{{van Dokkum}
  et~al.}{2018a}]{vanDokkum.etal.2018}
{van Dokkum} P.,  et~al., 2018a, \mn@doi [\nat] {10.1038/nature25767}, \href
  {https://ui.adsabs.harvard.edu/abs/2018Natur.555..629V} {555, 629}

\bibitem[\protect\citeauthoryear{{van Dokkum} et~al.,}{{van Dokkum}
  et~al.}{2018b}]{vanDokkum.etal.2018b}
{van Dokkum} P.,  et~al., 2018b, \mn@doi [\apjl] {10.3847/2041-8213/aab60b},
  \href {https://ui.adsabs.harvard.edu/abs/2018ApJ...856L..30V} {856, L30}

\bibitem[\protect\citeauthoryear{{van Dokkum}, {Danieli}, {Abraham}, {Conroy}
  \& {Romanowsky}}{{van Dokkum} et~al.}{2019a}]{vanDokkum.etal.2019}
{van Dokkum} P.,  {Danieli} S.,  {Abraham} R.,  {Conroy} C.,   {Romanowsky}
  A.~J.,  2019a, \mn@doi [\apjl] {10.3847/2041-8213/ab0d92}, \href
  {https://ui.adsabs.harvard.edu/abs/2019ApJ...874L...5V} {874, L5}

\bibitem[\protect\citeauthoryear{{van Dokkum} et~al.,}{{van Dokkum}
  et~al.}{2019b}]{vanDokkum.etal.2019b}
{van Dokkum} P.,  et~al., 2019b, \mn@doi [ApJ] {10.3847/1538-4357/ab2914},
  \href {https://ui.adsabs.harvard.edu/abs/2019ApJ...880...91V} {880, 91}

\bibitem[\protect\citeauthoryear{{van Dokkum} et~al.,}{{van Dokkum}
  et~al.}{2022a}]{vanDokkum.etal.2022}
{van Dokkum} P.,  et~al., 2022a, \mn@doi [\nat] {10.1038/s41586-022-04665-6},
  \href {https://ui.adsabs.harvard.edu/abs/2022Natur.605..435V} {605, 435}

\bibitem[\protect\citeauthoryear{{van Dokkum} et~al.,}{{van Dokkum}
  et~al.}{2022b}]{vanDokkum.etal.2022b}
{van Dokkum} P.,  et~al., 2022b, \mn@doi [\apjl] {10.3847/2041-8213/ac94d6},
  \href {https://ui.adsabs.harvard.edu/abs/2022ApJ...940L...9V} {940, L9}

\bibitem[\protect\citeauthoryear{{van der Burg}, {Muzzin}  \& {Hoekstra}}{{van
  der Burg} et~al.}{2016}]{van_der_Burg.etal.2016}
{van der Burg} R. F.~J.,  {Muzzin} A.,   {Hoekstra} H.,  2016, \mn@doi [\aap]
  {10.1051/0004-6361/201628222}, \href
  {https://ui.adsabs.harvard.edu/abs/2016A&A...590A..20V} {590, A20}

\bibitem[\protect\citeauthoryear{{van der Burg} et~al.,}{{van der Burg}
  et~al.}{2017}]{van_der_Burg.etal.2017}
{van der Burg} R. F.~J.,  et~al., 2017, \mn@doi [\aap]
  {10.1051/0004-6361/201731335}, \href
  {https://ui.adsabs.harvard.edu/abs/2017A&A...607A..79V} {607, A79}

\makeatother
\end{thebibliography}

\end{document}